\documentclass[aps,prx,superscriptaddress,longbibliography,twocolumn]{revtex4-2}

\pdfoutput=1

\usepackage{amssymb}
\usepackage{amsmath}
\usepackage{amsthm}
\usepackage{graphicx,bm}
\usepackage{epstopdf}
\usepackage{subfigure}
\usepackage{natbib}
\usepackage{epsfig}
\usepackage{amsfonts}
\usepackage{mathrsfs}
\usepackage{CJK}
\usepackage[toc,page,title,titletoc,header]{appendix}
\usepackage{array,longtable}
\usepackage{float}
\usepackage[usenames, dvipsnames, x11names]{xcolor}
\usepackage[colorlinks]{hyperref}
\usepackage[capitalise]{cleveref}
\usepackage{enumerate}
\usepackage{soul}
\usepackage{xcolor}
\usepackage{dsfont}
\hypersetup{
 colorlinks=true,
 citecolor=red,
 linkcolor=blue,
 urlcolor=cyan}
\usepackage{mathtools}
\usepackage{lipsum}

\newcommand{\mathcolorbox}[2]{\colorbox{#1}{$\displaystyle #2$}}

\newcommand{\tr}{\mathrm{tr}}
\newcommand{\sgn}{ \ \mathrm{sgn}}

\renewcommand{\L}{\hat{\mathcal{L}}}

\newcommand{\dd}{\ \mathrm{d}}

\newcommand{\D}{\mathcal{D}}
\newcommand{\U}{\hat{\mathcal{U}}}

\newcommand{\1}{\mathds{1}}

\newcommand{\SU}{\mathrm{SU}}

\renewcommand{\Re}{\mathrm{Re}}
\newcommand{\hc}{\mathrm{H.c.}}
\newcommand{\rss}{\hat \rho_\mathrm{ss}}
\newcommand{\E}{\mathds{E}}

\newcommand{\Erel}{\mathcal{E}_{2}}
\newcommand{\Gmax}{\Gamma_{\mathrm{max}}}
\newcommand{\kappab}{\overline{\kappa}}

\usepackage{comment}

\begin{document}

\title{Universal Time-Entanglement Trade-off in Open Quantum Systems}
\author{Andrew Pocklington}
\email{abpocklington@uchicago.edu}
\affiliation{Department of Physics, University of Chicago, 5640 South Ellis Avenue, Chicago, Illinois 60637, USA }

\affiliation{Pritzker School of Molecular Engineering, University of Chicago, Chicago, IL 60637, USA}

\author{Aashish A. Clerk}
\affiliation{Pritzker School of Molecular Engineering, University of Chicago, Chicago, IL 60637, USA}

\date{\today}

\begin{abstract}
We demonstrate a surprising connection between pure steady state entanglement and relaxation timescales in an extremely broad class of Markovian open systems, where two (possibly many-body) systems $A$ and $B$ interact locally with a common dissipative environment.  This setup also encompases a broad class of adaptive quantum dynamics based on continuous measurement and feedback.  As steady state entanglement increases, there is generically an emergent strong symmetry that leads to a dynamical slow down.  Using this we can prove rigorous bounds on relaxation times set by steady state entanglement.  We also find that this time must necessarily diverge for maximal entanglement.  To test our bound, we consider the dynamics of a random ensemble of local Lindbladians that support pure steady states, finding that the bound does an excellent job of predicting how the dissipative gap varies with the amount of entanglement.  Our work provides general insights into how dynamics and entanglement are connected in open systems, and has specific relevance to quantum reservoir engineering.  
\end{abstract}


\maketitle 


\section{Introduction}
\label{sec:0}

An exciting and powerful recent direction in quantum many-body physics is the realization that dynamical properties can be directly related to ground state entanglement features.  For example, pioneering work by Hastings showed that the ground states of finite-range, gapped 1D Hamiltonians obey an entanglement area law \cite{Hastings2007}. Under certain conditions, this result can be extended to longer-range interactions \cite{Kuwahara2020,Gong2017} and higher dimensions \cite{Masanes2009,Anshu2022,Brandao2015b,Cho2014,Michalakis2013,Wolf2008,deBeaudrap2010}, 
(c.f.~\cite{Eisert2010} for a review). By understanding aspects of a many-body system's energy spectrum, 
one can obtain extremely non-trivial information about the structure of entanglement in its ground state.

\begin{figure}[t!]
    \centering
    \includegraphics{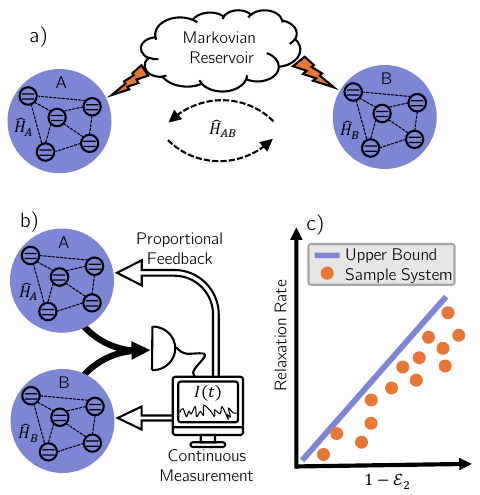}
    \caption{(a) Schematic showing two arbitrary systems $A$ and $B$ interacting dissipatively via local couplings to a common Markovian reservoir, along with arbitrary intra-system ($\hat H_{A}, \hat H_B$) as well as inter-system ($\hat H_{AB}$) Hamiltonian interactions.  We focus on cases where the dynamics leads to pure steady-states with finite entanglement. (b) The above dynamics is equivalent to a generic adaptive measurement setup: a continuous collective measurement is made of $A $and $B$, and the result $I(t)$ is used to drive both systems.  
    (c)  Our key result is that the relaxation rate to the steady state is bounded above by $1 - \Erel$, where $\Erel$ is the scaled steady-state entanglement [c.f.~\cref{eqn:ScaledEntanglement}]. 
    $(1 - \Erel)$ is zero for maximally entangled states, indicating the dissipative gap must close in such cases.  }
    \label{fig:1}
\end{figure}

In this work, we show that non-trivial connections between steady state bipartite entanglement and dynamical properties can also be established in open quantum systems supporting pure steady states. Focusing on many-body Markovian dissipative systems (described by a GKSL/Lindblad master equation \cite{Lindblad1976,Gorini1976}) satisfying only weak locality constraints, we show that systems with pure, {\it maximally} entangled steady states necessarily exhibit dynamical isolation: there is an emergent strong symmetry that makes it impossible to prepare the entangled steady state in finite time. This phenomenon implies a vanishing of the dissipative gap.  

Further, for systems with non-maximal steady state entanglement, we prove an 
inequality that sets a lower bound on the preparation time of the steady state.  This bound 
shows that the time required to reach the steady state grows exponentially in the Renyi-2 entanglement entropy of the steady state. The general setup we consider [c.f.~\cref{fig:1}] also directly constrains a broad class of measurement and feedback protocols, whose unconditional dynamics result in steady state entanglement. We thus establish a fundamental trade-off between steady state entanglement generation and relaxation times for an extremely wide class of non-unitary dynamics. Note that previous work on open systems has established relations between steady state correlations and the dissipative gap \cite{Kastoryano2013,Brandao2015,Poulin2010}; however, unlike our work, these results only apply in the thermodynamic limit, and do not connect bipartite entanglement and relaxation times.  


Our result has strong implications for the general techniques of reservoir engineering and autonomous feedback \cite{Poyatos1996, Plenio2002}.  Such approaches are ubiquitous in quantum information processing, and involve employing tailored dissipative processes to prepare and stabilize useful quantum steady states.  Perhaps the most common kind of target states here are those with long-range entanglement (see e.g.~\cite{Kraus2004,Schirmer2010,Stannigel2012,ZipilliPRL2013,Motzoi2016,Ma2017,Ma2019,Doucet2020,Zippilli2021,Agusti2022,Govia2022,Brown2022,Angeletti2023,
lingenfelter2023,Diehl2008,Kraus2008,Diehl2011,Pocklington2022,Zippilli2015,Ma2017}).
While extremely powerful, reservoir engineering is only practically effective if the relevant relaxation rates are sufficiently fast (otherwise intrinsic, uncontrolled dissipative processes will corrupt the eventual steady state).  Our fundamental entanglement-time trade-off directly applies to this setting. Previous work had phenomenologically seen evidence of such a trade-off in a variety of different schemes; i.e.,~the preparation time would diverge as the steady state was engineered to have maximal entanglement  \cite{Govia2022,Pocklington2022,Brown2022}. In the simplest, specific case of two qubits, this trade-off could be connected to an effective conservation of angular momentum that emerged in the maximum entanglement limit  \cite{Brown2022}.  Our results establish the origin of this trade-off in the most general many-body setting, and have far-reaching implications on the design of optimal entanglement stabilization protocols. 
Note that for two qubits, studies have shown how to evade dynamical slowdown by using additional energy levels \cite{Doucet2020,Brown2022}.  Such an approach effectively changes the local Hilbert space dimensions (and thus the maximum possible amount of entanglement), and thus a version of our bound still applies.


One might worry that while our work sets an upper bound on timescales for dissipative remote entanglement stabilization, this bound might be extremely loose, and have little relevance to typical systems. To address this, we first solve a seemingly complex inverse problem: given a specific many-body entangled pure state of interest, how do we reverse engineer a local dissipative process that will stabilize it?  We provide a very general construction that solves this challenge, and use it to construct a set of local, random Lindbladians that all stabilize a given target state.  Using this ensemble of random Lindbladians, we show that the dissipative gap scales with entanglement entropy exactly as predicted by our bound.  Moreover, we show analytically that the relaxation of a Haar random initial state follows the same scaling predicted by the bound.  Our general reverse engineering of local dissipation compatible with a target entangled state could have a variety of other interesting applications  \cite{Guo2024}. 

 One potential application of recent interest is designing open quantum systems which are able to simulate finite temperature states of many-body systems (often called quantum Gibbs samplers)\cite{Chen2023,Chen2023,Ding2024,Sun2024,Guo2024}.  Our technique could be used to prepare the purified thermofield double (TFD) state, which allows one to probe finite temperature properties. Another possible use is given in Ref.~\cite{Mamaev2024}, where our technique would be able to prepare an entangled state for optimal multi-parameter metrology.

This article is organized as follows.  \cref{sec:1} establishes our general setup, while \cref{sec:2} provides a rigorous statement of our main results. In \cref{sec:3}, we show how to reverse engineer a class of local dissipative dynamics that all stabilize a given target entangled state.  We combine this with random matrix theory to show that typical relaxation times scale
with entanglement entropy 
exactly as predicted by our bound.  In \cref{sec:4}, we show that the spectra of random Lindbladians exhibit a bulk gap along with isolated midgap state(s), which are responsible for all of the slow dynamics. 

\section{Setup and definitions}
\label{sec:1}

\subsection{Maximally Entangled States}
We work throughout with systems having a tensor product Hilbert space $\mathcal{H} = \mathcal{H}_A \otimes \mathcal{H}_B$
(with $\dim \mathcal{H}_{A,B} = N_{A,B} < \infty$), and will be interested in pure states with entanglement between subsystems $A$ and $B$.  
Unless otherwise stated, we will assume $N_A = N_B \equiv N$. Any state $|\psi \rangle \in \mathcal{H}$ admits a Schmidt decomposition
\begin{align}
    |\psi \rangle &= \sum_{i = 1}^{\min (N_A,N_B)} \sqrt{p_i} |i \rangle_A \otimes |i \rangle_B, \label{eqn:SchmidtDecomp}
\end{align}
where the Schmidt coefficients $\sqrt{p_i}$ are taken to be real and positive without loss of generality throughout. A \textit{maximally entangled state} for our bipartition is any state with uniform Schmidt coefficients, i.e.~$\sqrt{p_i} = [\min (N_A,N_B)]^{-1/2} \ \forall i$. 

The bipartite entanglement of $|\psi \rangle$ can be characterized by the Renyi-$\alpha$ entropy between subsystems $A$ and $B$:
\begin{align}
    S^{(\alpha)}(|\psi \rangle) &= \frac{1}{1 - \alpha} \log \sum_i p_i^{\alpha}. \label{eqn:Renyi}
\end{align}
For a maximally entangled state $|\psi \rangle_{\mathrm{max}}$ (letting $N_A = N_B = N$), this gives
\begin{align}
    S^{(\alpha)}(|\psi \rangle_{\mathrm{max}}) &= \log N.
\end{align}

\subsection{Open Markovian Dynamics and Timescales}

Our overarching goal is to connect steady state entanglement to dynamics.  
We will focus exclusively on systems whose open system dynamics is described by a Markovian master equation in Gorini-Kossakowski-Sudarshan-Lindblad (GKSL) form \cite{Gorini1976,Lindblad1976}.  Letting $\hat \rho$ be the system's density matrix, we have
\begin{align}
    \partial_t \hat \rho &= -i[\hat H, \hat \rho] + \sum_{\mu=1}^M \D[\hat L_\mu] \hat \rho \equiv \L \hat \rho, \label{eqn:me1} \\
    \D[\hat L_\mu] \hat \rho &\equiv \hat L_\mu \hat \rho \hat L_\mu^\dagger - \frac{1}{2} \left\{ \hat L_\mu^\dagger \hat L_\mu, \hat \rho \right\}, \label{eqn:me2}
\end{align}
where the Hermitian operator $\hat{H}$ is the Hamiltonian, and the $M$ jump operators $\hat L_\mu$ parameterize the non-unitary evolution.  We will refer to the superoperator $\L$ as the Lindbladian. 
As we discuss below, we consider situations where subsystems $A$ and $B$ are physically separated and only interact via local couplings to common dissipative environments.  As a result, we require all jump operators to have the form:  
\begin{align}
        \hat L_\mu &= \hat A_\mu \otimes \1 + \1 \otimes \hat B_\mu. \label{eqn:me3}
\end{align}

Every Lindbladian will have at least one steady state solution $\rss$ defined by $\L \rss = 0$.  Our interest here is in systems which have a pure steady state; the goal is to connect the entanglement of the steady state to dynamical timescales.  
We say that $\L$ has a maximally entangled steady state if there exists a state $|\psi \rangle$ such that $S^{(\alpha)}(|\psi \rangle) = \log N $ and $\L ( |\psi \rangle \langle \psi | ) = 0$. A steady state is unique if and only if every initial condition tends towards the steady state in the long time limit.

In most cases, we will be interested in systems where $\L$ is diagonalizable, and can be written as $\L = \sum_{\alpha = 0}^{N^4 - 1} \lambda_\alpha | r_\alpha \rangle \rangle \langle \langle l_\alpha |$. (Here and throughout, the double bracket notation $|\hat \rho \rangle \rangle$ represents a vectorized density matrix.) The complete set of dissipative rates are given by the real parts of the eigenvalues $\{ \lambda_\alpha \}$. We will work with the convention that $0 = \lambda_0 \geq \Re \lambda_1 \geq \dots \geq \Re \lambda_{N^4 - 1}$ \cite{ContractionMappingNote}. We define the dissipative gap 
\begin{align}
\Delta = -\Re \lambda_1. \label{eqn:diss_gap}
\end{align}

\subsection{Locality in Bipartite Lindbladians}

When proving an area law for the ground states of gapped 1D systems, 
one has to first
imbue the Hamiltonian with a meaningful notion of locality. We similarly must  identify a relevant notion of locality in our open system dynamics.  To that end, we assume that subsystems $A$ and $B$ are physically separated, and only interact via local couplings to common, extended Markovian reservoirs, see \cref{fig:1}. For example, one might consider groups of qubits decaying into a common waveguide, or groups of atoms interacting with a common cavity mode.  The locality of this setup means that before eliminating the environment to generate our master equation, its interaction with the system will be described by a Hamiltonian of the form:
\begin{align}
   \hat H_{\mathrm{int}} &= \sum_{\mu=1}^{\mathcolorbox{white}{M'}} \hat R_{A,\mu} \otimes \hat{\tilde{A}}_\mu \otimes \1 + \hat R_{B,\mu} \otimes \1 \otimes \hat{\tilde{B}}_\mu + \hc \label{eqn:sys-bath-coupling}
\end{align}
where $\mu$ indexes interactions with the reservoir(s), and $\hat R_{A,\mu}$ and $\hat R_{B,\mu}$ are reservoir operators localized near either the $A$ or the $B$ subsystem, respectively.  Correspondingly, $\hat{\tilde{A}}_\mu$ ($ \hat{\tilde{B}}_\mu)$ are subsystem $A$ ($B$) operators. Note crucially that $A$ and $B$ {\it only} interact via their common coupling to the environment.   

Assuming now the reservoir(s) is Markovian, we can eliminate it in the usual manner (see e.g.~\cite{Breuer2002,Gardiner2004}) to generate a GKSL master equation for the dynamics of $A+B$
with the constrained form of
\cref{eqn:me1,eqn:me2,eqn:me3}, see \cref{app:6} for more details. In particular, each jump operator $\hat{L}_\mu$ is the sum of an $A$ and a $B$ operator as given in  \cref{eqn:me3}, where $\hat A_\mu, \hat B_\mu$ are local $A,B$ operators  that depend on $\hat{\tilde{A}}_\mu$ and $ \hat{\tilde{B}}_\mu$ as well 
as reservoir properties. 
Even with this constraint, we have an extremely general problem.  We can still have arbitrary local dissipative processes 
(i.e.~set either $\hat A_\mu$ or $\hat B_\mu$ to $0$), as well as all forms of correlated Markovian dissipation relevant to physically separated systems.  Moreover, we place no constraints on the Hamiltonian $\hat{H}$ in \cref{eqn:me1} (as there could be arbitrary bath-induced Hamiltonian interactions between $A$ and $B$).  

\subsection{Connection to measurement-feedback dynamics}

The general setup described by \cref{eqn:me1,eqn:me2} is also directly relevant to describing dynamics where $A$ and $B$ interact via locality-constrained continuous measurement and feedforward (MFF) processes \cite{Wiseman1993,Wiseman1994,Metelman2017}.  In particular, this constrained form describes the unconditional dynamics arising from a MFF protocol where one measures {\it sums} of $A$ and $B$ quantities, and then uses the results to apply local feedback control to each subsytem.  To be explicit, consider the unconditional dynamics generated by making a weak continuous measurement of a Hermitian observable $\hat{M}$, and then using the measurement record to drive another Hermitian quantity $\hat{F}$.  In the limit where delay can be neglected, the theory of weak continuous measurement shows that the unconditional state (i.e.~averaged over all possible measurement outcomes) evolves as: 
\cite{Wiseman2009}
\begin{align}
    \partial_t \hat \rho &= \D[\hat M]\hat \rho + \D[\hat F] \hat \rho - i [\hat F, \{ \hat M, \hat \rho \}], \nonumber \\
    &= \D[\hat F - i \hat M] \hat \rho - \frac{i}{2} [\{ \hat F, \hat M \}, \hat \rho ].
    \label{eq:MFF}
\end{align}
As long as both $\hat{M}$ and $\hat{F}$ are sums of local operators, we have a master equation that obeys the general form of \cref{eqn:me1,eqn:me2,eqn:me3}.  For example, we could take:
\begin{align}    
    \hat M & = \frac{i}{2}(\hat A - \hat A^\dagger + \hat B - \hat B^\dagger) \\
    \hat F & = \frac{1}{2}(\hat A + \hat A^\dagger + \hat B + \hat B^\dagger)
\end{align}
In this case $\hat F - i \hat M = \hat A + \hat B$, implying the dissipator in \cref{eq:MFF} has the required form of \cref{eqn:me3} (while the last term in \cref{eq:MFF} is a Hamiltonian interaction which is always allowed). 

Given this connection, the entanglement-time bounds we prove below directly constrain locality-constrained measurement+feedback protocols. Moreover, we stress that while we have formulated the measurement+feedback protocol without any additional Hamiltonian interactions, any Hamiltonian can always be added in without any change to our results.

\section{Bound Statement}
\label{sec:2}

\subsection{Maximally Entangled Steady States Cannot be Reached by Markovian Dynamics}
\label{subsec:MaxENBound}

It is well known that dissipative dynamics having the form of 
\cref{eqn:me1,eqn:me2,eqn:me3} can be used to stabilize pure entangled states.  Examples include the dissipative stabilization of bosonic two-mode squeezed states \cite{Caves1985,Gerry1985,Pocklington2023}, qubit Bell pairs \cite{ZipilliPRL2013,Pocklington2022,Govia2022}, and even more exotic states of matter in spin chains \cite{Pocklington2022,lingenfelter2023}.  Our first  result is to show that all such protocols are highly constrained.  
If a Lindbladian $\L$ of the form \cref{eqn:me1,eqn:me2,eqn:me3} has a pure 
{\it maximally} entangled steady state $| \psi \rangle$
(i.e.~$\L |\psi \rangle \langle \psi | = 0$), then this state is necessarily dynamically isolated:  the projector $| \psi \rangle \langle \psi |$ becomes a conserved quantity, implying that the dissipative dynamics will never relax an arbitrary initial state into this entangled state.  This necessarily implies the existence of multiple steady states and the closing of the dissipative gap. Our result here holds irrespective of further details (e.g.~Hilbert space dimension, number of jump operators, form of the $\hat{A}_\mu, \hat{B}_{\mu}$ operators, form of $\hat{H}$, etc.).  

To establish this result, note first that as $|\psi \rangle$ is a pure steady state, we necessarily have \cite{Kraus2008}
\begin{align}
    [\hat H, | \psi \rangle \langle \psi |] 
    &= \hat L_\mu |\psi \rangle = 0,
\end{align}
i.e.,~it is an eigenstate of the Hamiltonian and a dark state of each jump operator.
Within the quantum jumps interpretation of our master equation \cite{Wiseman2009}, the dark state conditions imply that if the system is in the state $|\psi\rangle$, there is zero probability of a quantum jump evolving it into a different state.  

The fact that $|\psi\rangle$ is also maximally entangled leads to a second, even stronger constraint: there will also be zero probability that a quantum jump from an arbitrary initial state $|\phi \rangle$ will produce a state with non-zero overlap with $|\psi \rangle$. To see this explicitly, we define the unnormalized ``absorbing state'' associated with each jump operator $\hat{L}_\mu$ to be $|\tilde \psi_\mu \rangle \equiv \hat L_\mu^\dagger |\psi \rangle$. The probability that a quantum jump induced by $\hat{L}_\mu$ will result in some initial state $| \phi \rangle$ having overlap with $|\psi \rangle$ is then $|\langle \psi| \hat{L}_\mu | \phi \rangle|^2 = 
|\langle \tilde \psi_\mu | \phi \rangle|^2$. Using the fact that $|\psi \rangle$ is a dark state, we have:
\begin{align}
    & \langle \tilde \psi_\mu | \tilde \psi_\mu \rangle 
=  \langle \psi | \hat L_\mu \hat L_\mu^\dagger | \psi \rangle = \langle \psi | [\hat L_\mu, \hat L_\mu^\dagger] | \psi \rangle \nonumber \\
    &  \ \ = \langle \psi | [\hat A_\mu, \hat A_\mu^\dagger] \otimes \1  | \psi \rangle  + \langle \psi | \1 \otimes [\hat B_\mu, \hat B_\mu^\dagger] |  \psi \rangle.
    \label{eq:AbsorbingStateNorm}
\end{align}
Next, as $| \psi \rangle$ is also a maximally entangled state, an explicit calculation shows that the above expression is proportional to $\tr [\hat A_\mu, \hat A_\mu^\dagger] + \tr [\hat B_\mu, \hat B_\mu^\dagger] = 0$ when $N_A = N_B$ (see \cref{app:1}). Hence, there is zero probability for a quantum jump moving population into the entangled steady state $|\psi \rangle$. The vanishing of $| \tilde \psi_\mu \rangle$ also implies that the ``no-jump" evolution described by $\hat{H}_{\rm eff} = \hat{H} - (i/2) \sum_\mu \hat L_\mu^\dag \hat L_\mu$ will never increase the population of $| \psi \rangle$.  We thus have established our key result: the population of the maximally entangled steady state $| \psi \rangle$ will never change in time, and hence even given infinite time, dissipative preparation of the steady state $| \psi \rangle$ is impossible.  

We can also establish this result rigorously using the notion of a strong symmetry of a Lindbladian \cite{Buca2012}.  \cref{eq:AbsorbingStateNorm} shows that for each jump operator $\hat{L}^\dagger_\mu | \psi \rangle = 0$.  Given that $|\psi \rangle$ is also a pure steady state, it immediately follows that 
\begin{align}
    [\hat{P}, \hat L_\mu] = [\hat{P}, \hat H] = 0
    \label{eq:PStrongSymm}
\end{align}
where $\hat{P} = | \psi \rangle \langle \psi |$.
This implies that $\hat{P}$ (the projector onto $|\psi \rangle$) generates a strong symmetry of $\L$ and describes a dynamically conserved charge.  It thus separates the full Hilbert space into two dynamically isolated subspaces, namely $|\psi \rangle$ and its orthogonal complement.  It also tells us that there must be steady state degeneracy (i.e.~at least one steady state orthogonal to $| \psi \rangle$), and hence a vanishing of the dissipative gap.    
Our result here generalizes the discussion of \cite{Brown2022}, which discusses this phenomena in a specific two-qubit Lindbladian where the conserved quantity reduced to total angular momentum.    Our generalization shows that this phenomena occurs in an extremely broad class of systems (including systems in the truly many-body limit), and that the conserved quantity is the population of the entangled steady state itself.  

\subsection{Universal Time-Entanglement Trade-Off}

We now establish the second main result of this work:  a fundamental trade-off in our locality-constrained dissipative dynamics between the amount of pure steady state entanglement and the timescales associated with dissipative stabilization.  We find a rigorous bound that implies increased steady state entanglement leads to longer relaxation times (with the timescale diverging for maximal entanglement, as demonstrated above).  

To formulate these ideas, we again consider a Lindbladian $\L$ of the form in \cref{eqn:me1,eqn:me2,eqn:me3}, which has a pure stady state $\rss \equiv |\psi \rangle \langle \psi |$.  We now allow $\rss$ to have an arbitrary amount of entanglement.  Consider the time evolution of an arbitrary initial state under $\L$, with the time dependent density matrix denoted as $\hat\rho_t$.
We are interested in how this state relaxes towards $\rss$, and thus consider the fidelity $F(t)$ between these states:   
\begin{align}
    F(t) &= \left( \tr \sqrt{\sqrt{\rss} \hat \rho_t \sqrt{\rss}} \right)^2.
\end{align}
Relaxation of an initial trivial state to the entangled steady state corresponds to $F(t)$ evolving from $\sim 0$ at $t=0$ to $\sim 1$ at some finite time $t \sim \tau_{\rm rel}$.  

To formulate our result, consider the product basis defined by the Schmidt decomposition of our pure steady state given in \cref{eqn:SchmidtDecomp}. To fix a single time scale for the problem, we will non-dimensionalize the jump operators by pulling out an overall scale factor with units of frequency. This allows us to differentiate the role entanglement plays in the dynamics form the more trivial role the overall magnitude of the Lindbladian plays. I.e., we will define $|\hat A_\mu| \equiv \sqrt{\kappa_\mu}$ to be the magnitude of the largest matrix element of $\hat A_\mu$ in the Schmidt basis, so that we can write each jump operator as:
\begin{align}
    \hat L_\mu = \sqrt{\kappa_\mu} \left( \frac{\hat A_\mu}{|\hat A_\mu|} \otimes \1 + \1 \otimes \frac{\hat B_\mu}{|\hat A_\mu|} \right) \equiv \sqrt{\kappa_\mu} \hat{\tilde{L}}_\mu, \label{eqn:rescaledJump}
\end{align}
where $\hat{\tilde{L}}_\mu$ is unitless and $\kappa_\mu$ has the units of a rate. From here, we define the average rate $\overline{\kappa} = \frac{1}{M} \sum_{\mu = 1}^M \kappa_\mu$. We will go on to show that this average rate $\overline{\kappa}$ appears naturally in the models considered and sets an overall time scale.  Turning to the steady state $\rss = | \psi \rangle \langle \psi |$, we use $S^{(2)}_{\mathrm{ss}}$ to denote its Renyi-2 entanglement entropy [c.f. \cref{eqn:Renyi}], and define $\sqrt{p_\mathrm{min}}$ to be its smallest Schmidt coefficient. 

Finally, we introduce the scaled entanglement $\Erel$, a measure for the steady state entanglement based on $S^{(2)}$ that varies from $0$ (no entanglement) to $1$ (maximal entanglement):
\begin{align}
    \Erel \equiv \frac{1}{1-1/N} \left(1 - e^{-S^{(2)}_{\mathrm{ss}} } \right)
    = 
    \frac{1}{1-1/N} \left(1 - \sum_{j=1}^N p_j^2 \right)
    \label{eqn:ScaledEntanglement}
\end{align}

With these definitions in hand, we can state our key result:  the growth of the fidelity $F(t)$ is rigorously bounded by a rate, whose value is directly proportional to the entanglement deficit of the steady state.  Assuming first that $p_{\rm min} > 0$, we have:
\begin{subequations} \label{eqn:the_bound}
\begin{align}
    & |F(t) - F(0) |\leq  \Gmax t, 
    \label{eq:the_bound1}
    \\
    & \frac{\Gmax}{M \overline{\kappa}} = 
    \sqrt{2} \left(N-1\right) \left( p_{\mathrm{min}}\right)^{-1}    
   \, \left( 1 - \Erel \right), 
\end{align}
\end{subequations}
where $N \equiv N_A = N_B $.
A full proof of this result is presented in \cref{app:1}.
We see that the growth of the fidelity towards $1$ is bounded by the rate $\Gmax$, which in turn decreases linearly with the scaled entanglement $\Erel$.  For a maximally entangled state ($\Erel = 1$), $\Gmax$ vanishes, thus recovering the result of the previous subsection: $F(t)$ is time independent in this case, and no dynamical stabilization of the entangled steady state is possible.  For more general cases, our result provides a lower bound on the relaxation time 
$\tau_{\rm rel}$:  $\tau_{\rm rel} \geq 1/\Gmax \propto 1 / (1- \Erel)$.
At a heuristic level, for $\Erel < 1$ we do not have a perfect strong symmetry and conserved quantity like the maximal-entanglement case (c.f.~Eq.~(\ref{eq:PStrongSymm})).  Nonetheless, there is an ``almost" conserved quantity that relaxes slowly, leading to very slow relaxation to the steady state.  This is discussed in more detail in \cref{app:3}.

The bound in \cref{eqn:the_bound} is for the case where the steady state reduced density matrix of each subsystem is full rank; it clearly has no utility in the case where $p_{\rm min}$ is zero or extremely small.  In these cases, an analogous, more useful bound can be derived that again constrains relaxation timescales in terms of the steady state entanglement deficit.  The bound still has the form of \cref{eq:the_bound1}, but the rate $\Gmax$ is replaced by $\Gmax'$ (see \cref{app:1}):
\begin{align}
    \frac{\Gmax'}{\sqrt{2(N^3-N^2)} } & = 
        \left[ \left(\sum_{\mu=1}^M |\hat{A}_\mu| 
    \right)^2 + \left(
        \sum_{\mu=1}^M |\hat{B}_\mu|  \right)^2
    \right]
    \left(1 - \Erel \right)^{1/2} 
    \label{eq:ModBound}
\end{align}
The rate in this bound is still decreases monotonically with increasing scaled entanglement $\Erel$, and vanishes as one approaches maximal entanglement $\Erel = 1$. 

Finally, the bounds discussed here constrain the relaxation of {\it any} state towards the entangled pure steady state.  It thus sets a speed limit for even the optimal cases, where one has a fast-relaxing state.  It is interesting to instead ask about the relaxation of a {\it typical} state towards the entangled steady state.  We can also derive a general bound that applies to this situation.  Consider that at some time $t$ we have a Haar-random pure state of our system, $\hat \rho_t = \hat U | \phi_0 \rangle \langle \phi_0 | \hat U^\dagger$ where $\hat U$ is a Haar-random unitary, and $|\phi_0 \rangle$ is some arbitrary fixed state.  We can then derive a rigorous bound on the instantaneous change in the average fidelity $F(t)$ (see \cref{app:2})
\begin{align}
    \int_{\mathrm{Haar}} |\partial_t F| \dd \hat U  \leq  
    2 M \overline{\kappa} 
    \frac{N-1}{N^2} 
    \left( p_{\rm min} \right)^{-1}
    \left(1 - \Erel \right)
    \label{eqn:HaarBound}
\end{align}
We again find the same scaling with the scaled entanglement, but now with a smaller $N$-dependent prefactor.  

Finally, the same physics that leads to the bounds in \cref{eqn:the_bound,eq:ModBound,eqn:HaarBound} can also be used to bound the mixing time of the Lindbladian  \cite{temme2010}, a standard timescale metric for dissipative dynamics.  This involves using inequalities between quantum fidelity and the trace distance \cite{nielsen2001}. Explicitly, if we define 
\begin{align}
t_{\mathrm{mix}}(\epsilon) &= \inf \{ t > 0 \ | \ \forall \hat \rho, d_{\mathrm{tr}}(e^{\L t} \hat \rho, \hat \rho_{\mathrm{ss}} ) \leq \epsilon \},
\end{align}
where $d_{\mathrm{tr}}$ is the trace distance, then we show (see \cref{app:1}) that
\begin{align}
    t_{\mathrm{mix}}(\epsilon) \geq \frac{1 - \epsilon}{\Gmax} = \frac{(1 - \epsilon)M \overline{\kappa} p_{\mathrm{min}}}{\sqrt{2}(N-1)}(1 - \Erel)^{-1}, \label{eqn:MixingTime}
\end{align}
where $\Gmax$ is the rate introduced in \cref{eqn:the_bound}.

\cref{eqn:the_bound,eq:ModBound,eqn:HaarBound,eqn:MixingTime} are key results of this work.  They provide a unifying explanation for phenomena seen in specific studies of a variety of different dissipative systems, all of which observed an extreme slow down of dynamics as parameters were tuned to increase the entanglement of the dissipative steady state \cite{Govia2022,Pocklington2022,Brown2022}.
Moreover, while they have been formulated for 
the case where both systems have the same Hilbert space dimension $N_A = N_B$, they can be easily extended to the case where this is not true.  In this more general case, one can always map the problem to the case where dimensions are equal, but where one system is completely decoupled from some of its levels.  One could then directly apply the bound \cref{eq:ModBound}.  More importantly, such a setup is by definition never close to being maximally entangled by our definition (as it is not exploiting the full Hilbert space). See \cref{app:1} for more details.

\section{Many-Body Random Lindbladians}
\label{sec:3}

\begin{figure}[t!]
    \centering
    \includegraphics[width = \linewidth]{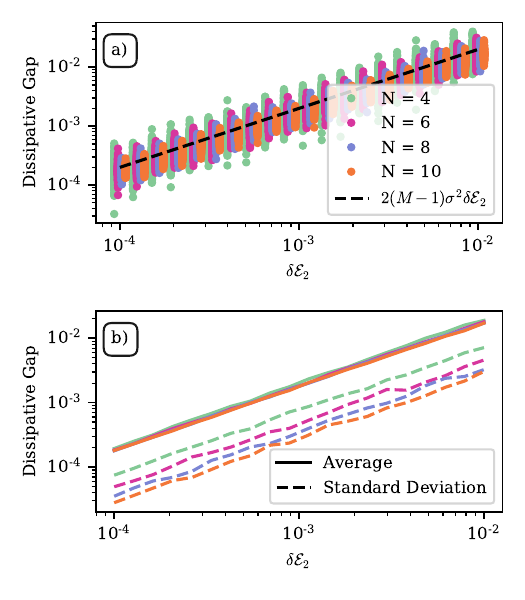}
    \caption{(a) Dissipative gap of random Lindbladians stabilizing a random pure state with a fixed entanglement $\delta \Erel$ [c.f. \cref{eqn:deltaE}], for varying system size constructed using \cref{eqn:A-B-Relation}.  Recall that $\delta \Erel$ decreases monotonically with increasing entanglement and is zero for a maximally entangled state.  Here, each $A_\mu$ matrix that is used to construct our dissipators is sampled from the Ginibre ensemble of random matrices with variance $\sigma$. For each system size and entanglement value, we plot results for 100 random Lindbladians. Different values of $N$ are offset horizontally to avoid overlap. (b) Average and standard deviation of data in (a) for a fixed system size and entanglement. The average gap is $2(M-1) \sigma^2 \delta \Erel$ and independent of system size. The standard deviation also scales linearly with $\delta \Erel$ and falls off as $N$ increases.
    }
    \label{fig:2}
\end{figure}

While our entanglement-time bound in \cref{eqn:the_bound} is rigorous, it only provides a lower bound on relaxation times.  There is no \textit{a priori} reason to assume this bound is tight, or that it even qualitatively captures how relaxation timescales vary in a typical system.  For example,  it could be that relaxation times diverge with increasing entanglement in a manner that is much worse than the predictions of our bound.  

To address these issues, in this section we study relaxation times in an ensemble of random Lindbladians having the form of \cref{eqn:me1,eqn:me2,eqn:me3}, that all have a pure steady state having some {\it fixed} value of the scaled entanglement $\Erel$ 
[c.f.~\cref{eqn:ScaledEntanglement}].
We can then ask about the statistics of relaxation times in this ensemble, and how they vary as we change the steady state entanglement.

\subsection{Reverse engineering dissipative dynamics compatible with a target entangled state}
\label{sec:Lindblad_Engineering}

To proceed, we consider a system with fixed local Hilbert space dimension $N$ for each subsystem, and start by picking a particular (perhaps randomly chosen) pure steady state $|\psi \rangle$ specified by its Schmidt coefficients $\sqrt{p_j}$ [c.f.~\cref{eqn:SchmidtDecomp}].  These coefficients can be used to define a single system operator $\hat \Psi$, which is nothing but the square root of the reduced steady state density matrix for each subsystem:
\begin{align}
    \hat \Psi  & \equiv \sum_i \sqrt{p_i} |i \rangle \langle i | 
    = \sqrt{\hat{\rho}_A} = \sqrt{\hat{\rho}_B}
    \label{eqn:BigPsi}
\end{align}
where $|i \rangle$ are states in the Schmidt basis for the entangled state $| \psi \rangle$, and  
$ \hat\rho_A = \tr_B |\psi \rangle \langle \psi |$, 
$ \hat \rho_B = \tr_A |\psi \rangle \langle \psi |$.  In what follows, we assume that all Schmidt coefficients are non-zero, i.e.~$\hat{\Psi}$ is full rank.  

The next step is to construct a dissipative dynamics of the form of \cref{eqn:me1,eqn:me2,eqn:me3} that has the chosen pure state $| \psi \rangle$ as a steady state.  We focus on the simple case where there is no Hamiltonian, and thus the problem reduces to finding one or more jump operators $\{ \hat L_\mu \}$ such that $\hat L_\mu |\psi \rangle = 0$, and where each jump operator is the sum of an operator acting on just one subsystem: $\hat{L}_\mu = \hat{A}_\mu + \hat{B}_\mu$.    Our approach is to first pick arbitrary system-$A$ operators $\hat{A}_\mu$ for each jump operator.  The dark state conditions then uniquely determine the form of the corresponding system $B$ operator in each $\hat{L}_{\mu}$.  
Letting $A_\mu, B_\mu$ and $\Psi_\mu$ denote the $N \times N$ matrix representation of these operators in the Schmidt basis, we have:
\begin{align}
    (\hat A_\mu \otimes \1 + \1 \otimes \hat B_\mu)|\psi \rangle &= 0 \implies B_\mu = -\Psi A^T_\mu \Psi^{-1}. 
    \label{eqn:A-B-Relation}
\end{align}

Our construction here provides an extremely general way to construct a large number of dissipative dynamics that will stabilize a particular chosen pure entangled steady state.
The construction guarantees that for a particular jump operator,  the operators $\hat A_\mu$ and $-\hat B_\mu$ are isospectral.  As a result, the kernel of a single jump operator $\hat L_\mu$ necessarily has a dimension $\geq N$, see \cref{app:1}.  Having a unique steady state will thus require at least two jump operators (chosen so that the desired steady state $|\psi\rangle$ spans the intersection of their kernels).  Alternatively, one could remedy this problem by introducing an appropriate Hamiltonian to the dynamics.  

We note that that the construction here
(where $B$ operators can be viewed as the modular conjugation of $A$ operators) has a close connection to certain formulations of quantum detailed balance (i.e.~KMS detailed balance \cite{Fagnola2007,Fagnola2010}
and hidden-time reversal symmetry \cite{Roberts2021}), as well as to the theory of coherent quantum absorbers \cite{Stannigel2012,Roberts2020}.  The construction is also reminiscent of the construction of the Petz recovery map \cite{Tsang2024}, and formal constructions of Hamiltonians that have thermofield double states as their ground state \cite{Cottrell2019}.

\subsection{Entanglement-time trade-offs and dissipative gap scaling in random many-body dissipative dynamics}

We now use our state-to-dynamics construction to assess whether the general time-entanglement bound in \cref{eqn:the_bound} tells us anything about typical relaxation times.  For a given chosen steady state entanglement $\Erel$, we first construct an ensemble of entangled pure states $|\psi_\alpha \rangle$ all having the same $\Erel$.  For each state, we then use our construction to generate a random Lindblad master equation having two jump operators that will stabilize this state. This involves first picking two random $N \times N$ matrices $A_1$, $A_2$, and then using \cref{eqn:A-B-Relation} to pick the corresponding $B$ matrices.  These matrices then define the jump operators $\hat{L}_1, \hat{L}_2$.  We draw each $A$ matrix from the complex random Ginibre ensemble:  each matrix element is a Gaussian random variable with $\E[(A_\mu)_{ij}] = 0$ and $\E[(A_\mu)_{ij}^*(A_\nu)_{kl}] = \sigma^2 \delta_{\mu \nu} \delta_{ik} \delta_{jl}$.  As we have no Hamiltonian contribution to our dynamics, the variance $\sigma^2$ plays no role except setting an overall timescale for the dynamics. That is to say, we can always choose to set $\sigma = 1$ by absorbing it into $\kappa_\mu$ as defined in \cref{eqn:rescaledJump}. In this way, we can separate the dimensionful quantity $\kappa_\mu$ from the dimensionless slowdown due to the entanglement entropy. Another way of putting it is to observe that, if we let $\L_\sigma$ be a Lindbladian sampled from the random ensemble where each matrix element of $A$ has variance $\sigma$, then $\L_\sigma = \sigma^2 \L_{\sigma = 1}$. Hence, if we normalize the Lindbladian to fix a time scale, the factors of $\sigma^2$ drop out, and can therefore be set to unity without loss of generality. Note that this is only true because there is no Hamiltonian, otherwise $\sigma$ would control the strength of the dissipative dynamics relative to the coherent dynamics.

In \cref{fig:2}, we show results obtained by numerically implementing this procedure for system sizes $N$ ranging from $4$ to $10$.
We plot the dissipative gap (c.f.~\cref{eqn:diss_gap}) for each constructed random Lindbladian, as a function of their steady state entanglement $\Erel$.  We stress that each realization here involves both randomly constructing a pure entangled steady state, and a dissipative dynamics that will stabilize this state.  The dissipative gap $\Delta$ characterizes the slowest relaxation process in our system, and hence we might expect that 
$\Delta \sim \Gmax$, where $\Gmax$ is the rate appearing in our general bound  \cref{eqn:the_bound}.
We see that there is a striking linear scaling of the average dissipative gap with $\Erel$, $\Delta \propto 1 - \Erel$.    
This is exactly the dependence predicted by our general bound for $\Gmax$.  While the prefactor of the scaling does not match the system-size dependence predicted by our bound, we see that the general trade-off between entanglement and relaxation times in this class of unstructured many-body Lindbladians is quantitatively in agreement with \cref{eqn:the_bound}.

Further, these results are not contingent on our use of the complex Ginibre ensemble to construct our random dissipators.  As shown in \cref{app:2}, constructions based on other, physically motivated random ensembles also show analogous scaling.   

Turning to the prefactor of the average dissipative gap scaling with $1-\Erel$, the bound states that it cannot grow faster than $N^2$. However, as shown in \cref{fig:2}(b), we numerically find that, for randomly sampled Lindbladians, it appears to be independent of $N$.  Further, we see that the fluctuations of the dissipative gap about its average decrease significantly even for modest increases in $N$. At this point, we stress that this does not imply that the bound is not tight, as it must account for \textit{all possible Lindbladians}, whereas in \cref{fig:2}, we consider only the more constrained set of \textit{average} or \textit{random Lindbladians}. In fact, in the next subsection, we will give a specific example of a system that relaxes much more quickly than the Haar random examples, but still satisfies the bound. To truly prove whether or not the bound is tight requires either finding a tighter bound, or finding a model system that saturates it, which we leave to future works.

However, we can analytically show that the Haar random case should have a prefactor that is independent of system size: first, define the deviation from maximal entanglement
\begin{align}
    \delta \Erel & \equiv \frac{N-1}{N}
    \left(1 - \Erel \right). \label{eqn:deltaE}
\end{align}
Then, assume that we have $M$ random dissipators constructed as above, and that for a fixed $\Erel$ the Schmidt coefficients $p_j$ are chosen randomly subject to normalization $\sum_j p_j = 1$ and $\Erel$ is fixed [c.f.~\cref{eqn:ScaledEntanglement}]. To leading order in $\delta \Erel$, one can show (see \cref{app:2}):
\begin{align}
    \E \left[ \int_{\mathrm{Haar}} |\partial_t F| \dd U \right]  = 2M \sigma^2 \delta \Erel, \label{eqn:Haar_Random_Bound}
\end{align}
where $\E[\bullet]$ is an average both over matrices $A_\mu$ as well as Schmidt coefficients $p_j$.  We now have a scaling of a typical relaxation rate that is still proportional to $(1-\Erel)$, but with an $N$-independent prefactor.

We expect that \cref{eqn:Haar_Random_Bound} will give a good estimate of the dissipative gap $\Delta$, except for the correction that $M \to M - 1$. The reason for this is that, due to the condition \cref{eqn:A-B-Relation}, a single jump operator generates $N$ steady states, and so the dissipative gap is by definition zero when $M = 1$. Despite this, the dynamics can still increase the fidelity to the chosen steady state, hence \cref{eqn:Haar_Random_Bound} is nonzero even when $M = 1$. To account for this difference between the dissipative gap and the rate of change of the fidelity, we expect that the average dissipative gap scales as
\begin{align}
\mathds{E}[\Delta] &= 2(M-1)\sigma^2 \delta \Erel. \label{eqn:HaarGap}
\end{align}
This is in good agreement with the behaviour of the average dissipative gap shown in \cref{fig:2}.

Note that for the Lindbladians considered here, we find that the dissipative gap accurately predicts long-time relaxation to the steady state.  It is well known that there exist examples where this correspondence can 
fail~\cite{Bensa2021,Bensa2022,Lee2023}, for example in systems exhibiting so-called ``skin-effects'' (see e.g.~\cite{Yao2018,Lee2016,Kunst2018,McDonald2018,Kawabata2019,Martinez2018,Haga2021}).  Note that even for cases where the dissipative gap is not reflective of relaxation, our general bound \cref{eqn:the_bound} remains valid:  it directly constrains the dynamics of the fidelity $F(t)$ (a physically observable quantity), without any assumptions on how the dissipative gap is related to the decay of observables.   

\subsection{Typical versus best case relaxation rates}

As noted above, it is unclear whether or not the bound stated in \cref{eqn:the_bound}is tight, as the random systems sampled from the given distribution seem to relax roughly $N^2$ times more slowly than the bound predicts is the fastest possible rate. However, we stress that there is nothing contradictory: the bound in \cref{eqn:the_bound} must account for all possible Lindbladians subject to the locality constraint, whereas in the random case, the systems are much less general.

For example, by adding back in more structure than is present in the randomly sampled Lindbladians, we can construct a model that, while not saturating the bound, does have a prefactor that is $\mathcal{O}(N)$, in between the one given in the bound and the Haar random case.

Consider a Lindbladian $\L_1$ acting on a Hilbert space with local dimension $d$, which has a steady state Renyi-2 entanglement entropy $S_1^{(2)}$ and a dissipative gap $\Delta_1$. We can enlarge the Hilbert space in a trivial way by tensoring together $n$ copies of the same steady state, which evolve under the Lindbladian

\begin{align}
    \L &= \sum_{i = 1}^n \1^{\otimes i-1} \otimes \L_1 \otimes \1^{\otimes n - i} \label{eqn:Structured_Lindbladian}
\end{align}
The steady state entanglement entropy (taking a bipartition that splits each copy) is now $S^{(2)}_{\mathrm{ss}} = nS_1^{(2)}$ by additivity. Thus, the deviation of the scaled entanglement from its maximum value scales as
\begin{align} 
    1 - \Erel &= \frac{N}{N-1} \left(e^{-S^{(2)}_\mathrm{ss} } - \frac{1}{N} \right) \\
    &= 
    \frac{1}{1-d^{-n}} \left(e^{-nS^{(2)}_1 } - d^{-n} \right),
\end{align}
which decays exponentially with $n$. 
It thus follows from \cref{eqn:Haar_Random_Bound}
that the instantaneous rate of change in fidelity between a Haar random state and the steady state is decaying {\it exponentially} in the number of copies $n$. However, 
by construction, the dissipative gap in this system is independent of $n$, suggesting no slow down with increasing system size.  This lack of slow down is however consistent with the larger prefactor in our more general bound in \cref{eqn:the_bound}.

This example shows that a general bound must have a prefactor that grows with $N$ at least as $\mathcal{O}(N/\log(N))$. The large deviation between this example and the Haar random models can be attributed to the highly structured form of the Lindbladian.  Stated succinctly, the eigenvectors of the Lindbladian are now very far from being related to Haar-random states.  All of the eigenvectors of \cref{eqn:Structured_Lindbladian}are completely unentangled between copies, and so approximating them with a Haar random state is unsuccessful.

\section{Beyond the slowest relaxation rate: structure of the Lindbladian spectra}
\label{sec:4}

\subsection{Bulk Gap and Midgap States}

The results of the previous section show that for random, unstructured dissipative dynamics that is locality-constrained and which has a pure steady state, the scaling of the dissipative gap with steady state entanglement matches the predictions of our general bounds in \cref{eqn:the_bound,eqn:HaarBound}.  In this section, we turn to another question:  does increasing steady state entanglement only lead to the formation of at most a handful of slow relaxation modes, or does it imply that an extensive number of relaxation modes become slow?  This is a question about the full spectrum of our Lindbladian, and not just the dissipative gap.  Through numerical investigation, we find that the first scenario holds:  strong steady state entanglement leads to the formation of a unique slow mode, whereas the vast majority of relaxation modes exhibit no slow down.  
We find that for a variety of different random Lindbladians, the spectrum of relaxation rates exhibit what we term a ``bulk gap'', where almost every mode has an $\mathcal{O}(\kappab)$ decay rate regardless of the slow down associated with large steady state entanglement. However, there also always exists a single, isolated ``midgap'' state, which decays extremely slowly and is responsible for all of the long-time slow dynamics.

To provide context for this result, we first recall known results for the spectral properties of completely unstructured random Lindbladians (i.e.~dynamics that do not have the locality and purity constraints of our general setting).  
Consider a general Lindbladian for a system with an $N$ dimensional Hilbert space:
\begin{align}
    \L \hat \rho &= \sum_{\mu, \nu=1}^N K_{\mu \nu} \left(  \hat F_{\mu} \hat \rho \hat F_\nu^\dagger - \frac{1}{2} \left\{ F_\nu^\dagger \hat F_\mu, \hat \rho \right\} \right),
\end{align}
where $K_{\mu \nu}$ is the complex positive semi-definite $N \times N$ ``Kossakowski matrix''. We take $K = NGG^\dagger/\tr(GG^\dagger)$, with the matrix $G$ sampled from the complex Ginibre ensemble (unit variance of matrix elements).  The operators 
$\{ \hat F_\mu \}$ form an orthonormal traceless basis of $\SU(N)$. 

For this general setup, it can be shown that the average dissipative gap scales as $1 - 2/N$ \cite{Denisov2019}.  
For large $N$ the average gap becomes $N$-independent, a scaling that will match almost all relaxation modes in our more structured dissipative problem.  However, the additional constraints we impose in our general setup (entangled pure steady state, local form of dissipators) leads to the formation of a single extremely slow mode as entanglement is increased, i.e. a ``midgap state", see \cref{fig:3_1}(a).  It is this single slow mode that is responsible for the slow relaxation described by our bounds. It is also interesting to note that the midgap state(s) are purely a result of the entanglement and locality constraints, they never show up in the completely random Lindbladians considered in Ref.~\cite{Denisov2019}.

Heuristically, this behaviour matches our general picture [c.f.~\cref{eq:PStrongSymm}] that as steady state entanglement increases, we have the emergence of an almost-conserved quantity, the projector onto the steady state.  This separates the Hilbert space into two subspaces.  One naively expects fast dynamics within each of these subspaces, with a single slow rate corresponding to mixing between the subspaces (see \cref{app:3} for more details).  This picture matches the results of our numerics.      
One can picture the steady state and midgap state(s) as forming a quasi-stable, slowly relaxing ``slow manifold'' whereas the rapidly decaying modes above the bulk gap are an effective ``fast manifold''. One could imagine tracing out the rapidly decaying states above the bulk gap to understand the slow dynamics within the midgap states(s); see, e.g. Ref.~\cite{Reiter2012}.
It is especially interesting to note, though, that in this case the slow manifold contains only a handful of modes, whereas the fast manifold is growing exponentially with $N$.

\subsection{Prethermalization and local 
relaxation}

Given this hierarchy of relaxation rates, it is interesting to ask what kind of observables relax slowly via the midgap state, and which relax quickly. Previous work on specific non-random models suggests that observables local to one subsystem tend to relax on fast $\mathcal{O}(\kappab)$ time scales, whereas non-local intersystem correlations relax slowly (see e.g.~\cite{Pocklington2022}). 

\begin{figure}[t!]
    \centering
    \includegraphics[width = 2.9in]{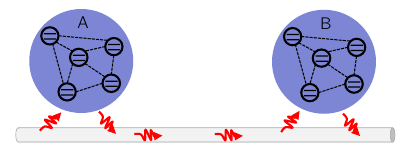}
    \caption{Schematic showing two systems $A$ and $B$ 
    that are coupled via a Markovian chiral (i.e.~directional) waveguide, thus realizing a cascaded quantum system.  
    The $A$ system is ``upstream'', and its dynamics are unaffected by $B$ (whereas $B$ is indeed affected by $A$).}
    \label{fig:CQA_Schematic}
\end{figure}

In order to study this, we want a system where we can independently look at the relaxation rate of the full $A-B$ system, as well as trace out one subsystem and look at the spectra of the other subsystem, which controls local observables. This can be achieved by constructing a unidirectional or cascaded \cite{Carmichael1993,Gardiner1993,Metelmann2015} quantum system following the so-called ``Coherent Quantum Absorber'' (CQA) approach \cite{Stannigel2012}, see \cref{fig:CQA_Schematic}.
Such a system still has the basic struture of our generic setup, where dissipative interactions between $A$ and $B$ are generated from local couplings to a common bath.  Now, however, these interactions are directional: $A$ influences $B$, but not vice-versa.

To achieve this directionality, and still have the dynamics stabilize a chosen pure entangled steady state, we chose the matrices $A_\mu$ [c.f. \cref{eqn:A-B-Relation}] in each jump operator so that they obey the constraint equation:
\begin{align}
    A_\mu = \Psi A_\mu^T \Psi^{-1}, \label{eqn:det_bal}
\end{align}
where $\Psi$ is defined via the steady state as per \cref{eqn:BigPsi}.
This condition corresponds to imposing an effective classical detailed balance condition (see \cref{app:4}).
We also add a Hamiltonian to our system $\hat H_\mathrm{CQA}$ that combines with each dissipator to enforce directionality, see \cref{app:4}. 
Note that previously, we used \cref{eqn:A-B-Relation} to define a dissipator given an arbitrary matrix $A_\mu$; now, \cref{eqn:det_bal} also gives a constraint on which $A_\mu$ are allowed. Using both \cref{eqn:A-B-Relation,eqn:det_bal} also tells us that $A_\mu = -B_\mu$, thus determining the form of each jump operator.
Using this construction, we find that the dynamics of system $A$ is given by
\begin{align}
    \tr_B [ \partial_t \hat \rho] &= \tr_B \left[ -i[\hat H_\mathrm{CQA}, \hat \rho] + \sum_\mu \D[\hat A_\mu + \hat B_\mu] \hat \rho \right] \nonumber \\
    &= -i[\hat H_A, \hat \rho_A] + \sum_\mu \D[\hat A_\mu] \hat \rho_A \equiv \L_A \hat \rho_A,
\end{align}
where $\hat \rho_A \equiv \tr_B \hat \rho$. It thus follows that every eigenvalue of the system $A$ Lindbladian $\L_A$ is also an eigenvalue of the full Lindbladian $\L$. Moreover, all observables on the $A$ system relax according to the spectrum of $\L_A$, whereas correlations between the two systems relax on a timescale governed by the gap of $\L$.  Note that because of \cref{eqn:det_bal}, the Lindbladian $\L_A$ necessarily satisfies classical detailed balance (also known as GNS detailed balance) \cite{Roberts2021,Tarnowski2023}.

\begin{figure}[t!]
    \centering
\includegraphics[width = \linewidth]{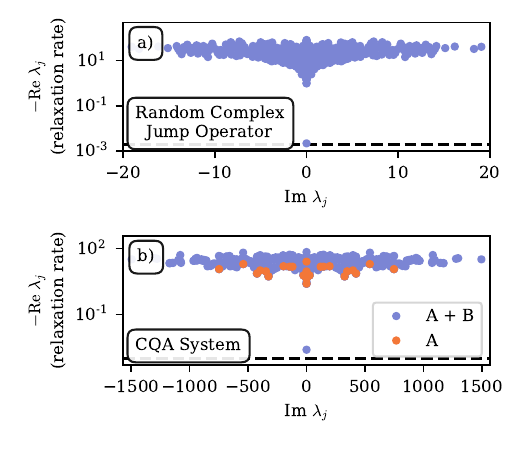}
\caption{
(a) Spectrum of a realization of a random Lindbladian on a bipartite system, using the construction of \cref{sec:Lindblad_Engineering}. Each jump operator is constructed by drawing the matrix 
$A_\mu$ from the complex Ginibre ensemble with variance $\sigma$; this then determines $B_\mu$ via \cref{eqn:A-B-Relation}.  
(b) Spectrum of a realization of a random Lindbladian constructed so that the interaction between $A$ and $B$ is necessarily directional (see \cref{fig:CQA_Schematic} and \cref{eqn:det_bal}). Blue points show the spectrum of the full Lindbladian $\L$, and orange points of the Lindbladian $\L_A$ describe system-$A$ only dynamics.  When looking at $\L_A$ alone, there is no emergent slow mode, and the dissipative gap is large.  
In both (a) and (b) the local dimension $N=5$, there are $M=2$ independent jump operators, and a pure, random steady state with fixed $\delta \Erel \equiv 10^{-3}$ [c.f.~\cref{eqn:deltaE}]. In both (a) and (b) the black dashed line shows $2(M-1) \sigma^2 \delta \Erel$ [c.f.~\cref{eqn:Haar_Random_Bound}], and the steady state ($\lambda_0 = 0$) is not shown.
}
\label{fig:3_1}
\end{figure}

In \cref{fig:3_1}(b), we numerically sample a random distribution of Schmidt coefficients $\{p_i\}$, and constrain $A_\mu$ to obey the detailed balance condition \cref{eqn:det_bal}. We also add the requisite Hamiltonian, making the system completely directional. Numerically, we observe that the full Lindbladian acting on both $A$ and $B$ has a small dissipative gap $\Delta \sim \kappab \delta \Erel$, as predicted by the bound and consistent with a random jump operator as in \cref{fig:3_1}(a). However, 
if we instead consider the spectrum of $\L_A$, the Lindbladian of system $A$ alone [c.f.~\cref{fig:CQA_Schematic}], the dissipative spectrum exhibits a large dissipative gap $\Delta \sim \kappab$.  This is in line with the results of 
Ref.~\cite{Tarnowski2023}, which studies the spectral properties of
random Lindbladians satisfying classical (GNS) detailed balance.  
This separates the dynamics into two regimes. The first is a ``prethermal'' regime characterized by the $\mathcal{O}(\kappab)$ bulk gap; during this time, local observables can relax to their steady state values as evidenced by the upstream system not experiencing slowdown. However, intersystem correlations and entanglement approach their steady state values on exponentially longer time scales characterized by the true dissipative gap, see \cref{fig:3_1}. This is in line with previously observed open system dynamics with highly entangled steady states (that are not necessarily directional), see e.g. \cite{Pocklington2022}.

\subsection{Multiple slow modes}

Our discussion of Lindbladian spectra has so far focused on cases where, apart from our constraints on locality and having a pure entangled stead state, the dynamics is essentially unstructured.  The slow-down of dynamics associated with increasing entanglement in this case can be attributed to a the emergence of a single slow mode.  We now ask how this situation is modified when our Lindbladian has some additional structure.  We find regimes where now multiple slow modes (midgap states) arise due to increasing steady state entanglement.  

Consider the case where we also have a notion of spatial locality {\it within} both the $A$ and $B$ subsystems. For example, consider two $n$ qubit spin chains denoted $A$ and $B$, with local XXZ Hamiltonians governed by the master equation $\partial_t \hat \rho = -i[\hat H_{\mathrm{XXZ}}, \hat \rho] + \L_{\mathrm{diss}}$ with 
\begin{align}
    \hat H_{\mathrm{XXZ}} &=  \sum_{s = A,B} \sum_{i = 1}^{n-1}  J \left( \hat \sigma^+_{s,i} \hat \sigma^-_{s,i + 1} + \hc \right) \nonumber \\
    & \ \ \ \ + \sum_{s = A,B} \sgn(s) \sum_{i = 1}^{n-1}  J_z \hat \sigma^z_{s,i} \hat \sigma^z_{s,i + 1}, \label{eqn:XXZs} \\
    \L_\mathrm{diss} &= \D[u\hat \sigma_{A,1}^- + v \hat \sigma_{B,1}^+] + \D[u\hat \sigma_{B,1}^- + v\hat  \sigma_{A,1}^+]. \label{eqn:local_jump}
\end{align}
Here, we take $\sgn(A) = -\sgn(B) = 1$ and $u^2 + v^2 = 1$.
This models two spins chains being driven by two-mode squeezed vacuum light at their boundary, and in the limit $J_z = 0$ has been considered as a method of entanglement generation \cite{ZipilliPRL2013,Pocklington2022,Angeletti2023,lingenfelter2023}.
This system has a pure steady state, and thus is an example of the general class of dynamics [c.f.~\cref{eqn:me1,eqn:me2,eqn:me3}] that we consider.  The steady state can be found exactly (c.f. \cite{lingenfelter2023,Pocklington2022} for related models), and is independent of $J,J_z$:
\begin{align}
    |\psi \rangle_\mathrm{ss} = \bigotimes_{i = 1}^n \left[\sqrt{1-v^2}|0\rangle_{A,i} |0\rangle_{B,i} + (-1)^i v|1\rangle_{A,i} |1 \rangle_{B,i} \right]. \label{eqn:rainbow_steadystate}
\end{align}
It follows that the steady state entanglement is controlled by $v$. 

\begin{figure}[t]
    \centering
\includegraphics[width = \linewidth]{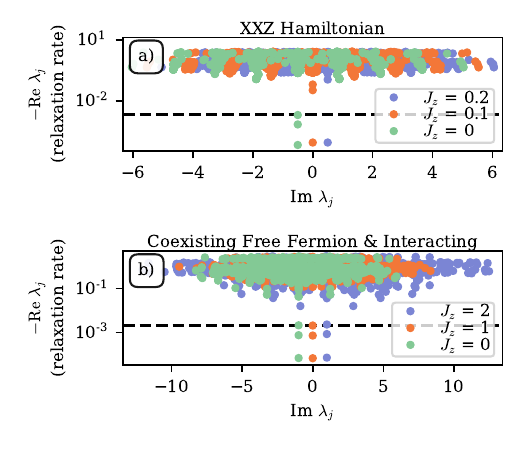}
\caption{
(a) Lindbladian spectra for the master equation for two $n=3$ qubit spin chains given by the XXZ Hamiltonian \cref{eqn:XXZs} and local boundary dissipation \cref{eqn:local_jump} for varying values of the intrachain ZZ interaction $J_z$, with $J=1$. Note whenever $J_z > 0$ (and the Hamiltonian is no longer mappable to free fermions), all but one of the midgap states move into the bulk spectra.
(b) Lindbladian spectra for the master equation for two $n=3$ qubit spin chains given by $\hat H = \hat H_\parallel + \hat H_\perp$ [c.f.~\cref{eqn:Hpar,eqn:Hperp}] and local boundary dissipation \cref{eqn:local_jump} for varying values of the interchain ZZ interaction $J_z$. Note this Hamiltonian always supports ballistic dynamics and the slow modes are robust to interactions.
In both (a) and (b) the steady state entanglement is fixed to be $\delta \Erel = 10^{-3}$ and the black dashed line is at $2(M-1)\delta \Erel$. Different values of $J_z$ are offset horizontally to avoid overlap. 
}
\label{fig:3_2}
\end{figure}

This is system is clearly more structured than the completely random examples studied in the previous subsections.  As such, one might expect that the Lindbladian spectrum would be very different, with potentially many more slow modes emerging when the steady state has high entanglement.  Surprisingly, for generic parameters this is not the case, see \cref{fig:3_2}:  one still gets a single slow mode.  

However, the special case where $J_z = 0$
the situation is very different.  For this parameter choice the local Hamiltonian \cref{eqn:XXZs} is equivalent to a free fermion Hamiltonian, and we find the emergence of {\it multiple} slow modes for strong entanglement, see \cref{fig:3_2}. Note that while the Hamiltonian alone can be mapped to non-interacting particles, the full dissipative dynamics still corresponds to an interacting fermionic problem (see \cite{Pocklington2022} for more details). As such, it is surprising at first glance that the non-interacting nature of the Hamiltonian alone leads to such differences in the dissipative spectrum.   Moreover, we observe numerically that the number of slow modes is exactly equivalent to $n$, the number of qubits in each chain, and hence is extensive in system size.

To see that the free fermion dynamics is truly the cause of having multiple slow modes, we can consider a more complicated local Hamiltonian introduced in Ref. \cite{Znidaric2013} which exhibits both free-fermion sectors of the Hilbert space, as well as diffusive sectors, see \cref{app:5}. Specifically, consider the master equation $\partial_t \hat \rho = -i[\hat H, \hat \rho] + \L_{\mathrm{diss}}$ with $\L_{\mathrm{diss}}$ defined as in \cref{eqn:local_jump} and $\hat H = \hat H_{\parallel} + \hat H_{\perp}$ with
\begin{align}
    \hat H_{\parallel} &= J\sum_{i = 1}^{n-1} \sum_{s = A,B} \hat \sigma_{s,i}^+ \hat \sigma_{s, i + 1}^- + \hc, \label{eqn:Hpar} \\
    \hat H_{\perp} &= \sum_{i = 1}^{n} J \left( \hat \sigma^x_{A,i} \hat \sigma^x_{B,i} + \hat \sigma^y_{A,i} \hat \sigma^y_{B,i} \right) + J_z \hat \sigma^z_{A,i} \hat \sigma^z_{B,i}. \label{eqn:Hperp}
\end{align}
Here, $\hat H_\parallel$ is equivalent to the XXZ Hamiltonian in \cref{eqn:XXZs} at the free fermion point $J_z = 0$. $\hat H_\perp$ couples the two chains together to form a two-rung ladder.  Recall that a direct Hamiltonian coupling between subsystems $A$ and $B$ is allowed by our general locality constraint, and does not impact the validity of our time-entanglement bounds.  One can again show that the steady state is pure and given by \cref{eqn:rainbow_steadystate}. Moreover, by plotting the dissipative spectra, we observe that there are still $n$ slow modes separate from the bulk spectra, just as in the completely free case, see \cref{fig:3_2}.

\section{Conclusion}
\label{sec:7}

In this paper, we have established a set of relations between pure steady state entanglement and relaxation dynamics for a class of many-body open quantum systems, where two systems $A$ and $B$ have locality-constrained dissipative interactions (i.e.~all dissipators are sums of local operators).  We find that such a system can never have a unique, maximally entangled steady state.  Further, we demonstrated that this result is a special case of a more general bound, which says that the time to reach the steady state is bounded below by how close that state is to being maximally entangled.  

We further explored this bound in the context of random Lindbladians satisfying our constraints, demonstrating that our bound accurately predicted the scaling of the dissipative gap with the steady state entanglement.  We also considered the Lindbladian spectra of such models, finding that for large entanglement, they generically have a bulk gap accompanied by extremely slow midgap state(s), the number of which we conjecture is related to whether or not the Hamiltonian is mappable to free fermions.

We believe that these results provide new insight into dissipative entanglement generation. They are directly relevant to quantum reservoir engineering schemes targeting remote entanglement,  and helps explain a number of previous results that observed a trade-off between entanglement and preparation time in various specific systems. In future work, it would be interesting to explore further whether such entanglement-time constraints also apply to more general situations (e.g.~extended many body systems where one could try to connect entanglement and relaxation times for different choices of regions $A$ and $B$).  

\section*{Acknowledgements}

This work was supported by the Air Force Office of Scientific Research MURI program under Grant No. FA9550-19-1-0399, the National Science Foundation QLCI HQAN (NSF Award No. 2016136), the Army Research Office under Grant No. W911NF-23-1-0077, and from the Simons Foundation through a Simons Investigator Award (Grant No. 669487). This work was completed in part with resources provided by the University of Chicago’s Research Computing Center. 

\bibliography{ref}

\begin{thebibliography}{80}
\bibitem [{Con()}]{ContractionMappingNote}%
  \BibitemOpen
  \href@noop {} {}\bibinfo {note} {Observe that because the (e.g. Bures) metric
  is contractive under completely positive, trace preserving maps, then by the
  contraction mapping theorem the Lindblad generator acting on a finite
  dimensional Hilbert space must have a zero eigenvalue. Hence, we define this
  zero eigenvalue to be $\lambda_0$, and arrange all remaining eigenvalues
  according to their real parts.}\BibitemShut {Stop}%
\end{thebibliography}

\appendix

\section{Bound Proof}
\label{app:1}

\subsection{Trace Relation of Maximally Entangled States}
Here, we will prove the trace property of maximally entangled states mentioned in the main text. Namely, if we define
\begin{align}
    |\psi \rangle &= \frac{1}{\sqrt{N}}\sum_{i =1}^N |i \rangle_A \otimes |i\rangle_B,
\end{align}
then for any local operator $\hat O_A \otimes \1$, we can see that
\begin{align}
    \langle \psi | \hat O_A \otimes \1 | \psi \rangle &= \frac{1}{N} \sum_{i,j} \langle i | \hat O_A |j \rangle \otimes \langle i | j \rangle \nonumber \\
    &= \frac{1}{N} \sum_{i} \langle i | \hat O_A |i \rangle = \frac{1}{N} \tr \hat O_A.
\end{align}
By the symmetry of the state under $A \leftrightarrow B$, this also tells us that the expectation value of any local $B$ operator of the form $\1 \otimes \hat O_B$ is also equivalent to its trace. Since for any jump operator $\hat L = \hat A \otimes \1 + \1 \otimes \hat B$ that is a sum of local operators, it's commutator with its adjoint is also a sum of local operators
\begin{align}
    [\hat L, \hat L^\dagger] &= [\hat A, \hat A^\dagger] \otimes \1 + \1 \otimes [\hat B, \hat B^\dagger],
\end{align}
then the expectation value 
\begin{align}
    \langle \psi | [\hat L, \hat L^\dagger] | \psi \rangle &= \frac{1}{N} \left( \tr  [\hat A, \hat A^\dagger] + \tr  [\hat B, \hat B^\dagger] \right) = 0,
\end{align}
as noted in the main text.

\subsection{Proof of Main Bound}
We can now move on to proving the main bound. Following a similar construction to \cite{delCampo2013}, we will bound the fidelity to the steady state 
\begin{align}
    F(t) &= \left( \tr \sqrt{\sqrt{\rss} \hat \rho_t \sqrt{\rss}} \right)^2.
\end{align}
Because we assume the steady state is pure, this can be simplified as
\begin{align}
    F(t) &= \tr (\hat \rho_t \rss).
\end{align}
Now, we can bound the change in the fidelity by its maximal derivative, i.e.
\begin{align}
    |F(t) - F(0)| \leq \Gmax t, \\
    \Gmax = \max |\partial_t F(t)|,
\end{align}
and so we will focus now on bounding $\partial_t F(t)$. Letting $\L$ be the 
Lindbladian and noting that $\rss$ is time-independent, we can observe that
\begin{align}
    |\partial_t F(t)| &= |\tr (\rss \partial_t \hat \rho_t)| = | \tr(\rss \L \hat \rho_t) | \nonumber \\
    &= | \tr(\L^\dagger \rss \hat \rho_t) |, \label{eqn:intermediate_result_2}
\end{align}
where we have used the definition of the adjoint Lindbladian to move it onto steady state (where the adjoint is with respect to the Hilbert-Schmidt norm). The Cauchy-Schwartz inequality gives:
\begin{align}
    \Gmax &\leq | \tr(\L^\dagger \rss \hat \rho_t) | \leq \sqrt{\tr(\L^\dagger \rss )^2 \tr(\hat \rho_t)^2} \nonumber \\
    &\leq \sqrt{\tr(\L^\dagger \rss )^2}.
\end{align}
Thus the rate $\Gmax$ is bounded by the Hilbert-Schmidt norm of the adjoint Lindbladian acting on the steady state. We can further simplify this:
\begin{align}
    \L^\dagger \rss &= i[\hat H, \rss] + \sum_\mu \hat L_\mu^\dagger \rss \hat L_\mu - \frac{1}{2} \{ \hat L_\mu^\dagger \hat L_\mu, \rss \} \nonumber \\
    &= \sum_\mu \hat L_\mu^\dagger \rss \hat L_\mu; \\
    \implies &\tr (\L^\dagger \rss)^2 = \tr \left[ \sum_{\mu,\nu} \hat L_\mu^\dagger \rss \hat L_\mu \hat L_\nu^\dagger \rss \hat L_\nu \right] \nonumber\\
    &= \sum_{\mu,\nu} |\langle \psi | [\hat L_\mu, \hat L_\nu^\dagger] | \psi \rangle |^2, \label{eqn:A1}
\end{align}
where we have used repeatedly the fact that $[\hat H, \rss] = \hat L_\mu \rss = 0$. Next, we will expand out the jump operator in terms of its matrix elements in the Schmidt basis:
\begin{align}
    \hat L_\mu &= \hat A_\mu \otimes \1 + \1 \otimes \hat B_\mu \nonumber \\
    &\equiv \sum_{ij}^N (A_\mu)_{ij} |i\rangle \langle j | \otimes \1 + \1 \otimes (B_\mu)_{ij} |i \rangle \langle j |.
\end{align}
To proceed, we will need to understand the relation between $ (A_\mu)_{ij}$ and $(B_\mu)_{ij}$ which is implied by $\hat L_\mu |\psi \rangle = 0$. We find that, writing $|\psi \rangle$ in the Schmidt basis, (and removing tensor product signs for brevity)
\begin{align}
    0 &= \sum_{ijkl}^N \sqrt{p_l} \bigg[  (A_\mu)_{ij} |i k\rangle \langle j k | + (B_\mu)_{ij} |k i \rangle \langle k j | \bigg]  |l l \rangle \nonumber \\
    &= \sum_{ij}^N  \bigg[ \sqrt{p_j} (A_\mu)_{ij}  + \sqrt{p_i} (B_\mu)_{ji} \bigg] |ij \rangle; \\
    \implies& (A_\mu)_{ij} = -\sqrt{p_i} (B_\mu)_{ji} \frac{1}{\sqrt{p_j}}.
\end{align}
Recalling the definition of $\Psi$ in \cref{eqn:BigPsi}, this is equivalent to 
\begin{align}
    A_\mu = - \Psi B^T_\mu \Psi^{-1},
\end{align}
as stated in the main text [\cref{eqn:A-B-Relation}]. Coming back to the bound, we can expand \cref{eqn:A1} in terms of matrix elements as
\begin{widetext}
\begin{subequations}
    \label{eqn:A2}
\begin{align}
    \Gmax^2 &\leq \sum_{\mu,\nu} \left[  \sum_{j,k = 1}^N  \left( A_\mu)_{j,k} (A_\nu)_{j,k}^*   + (B_\mu)_{j,k} (B_\nu)_{j,k}^* \right) \left(  p_j -  p_k \right) \right]^2   \\
    &= \sum_{\mu,\nu} \left[  \sum_{j,k = 1}^N  ( A_\mu)_{j,k} (A_\nu)_{j,k}^* \frac{\left(  p_j -  p_k \right)^2}{p_k} \right]^2 
    \leq  \sum_{\mu,\nu}  \left[ \frac{|A_\mu| |A_\nu|}{p_\mathrm{min}}  \sum_{j,k = 1}^N  \left(  p_j -  p_k \right)^2 \right]^2 \label{eqn:bound1} \\
    &= \frac{1}{p_\mathrm{min}^2}\left( \sum_{\mu} |A_\mu|^2 \right)^2 \left[  \sum_{j,k = 1}^N  \left(  p_j - p_k \right)^2 \right]^2 = \frac{2N^2}{p_\mathrm{min}^2}\left( \sum_{\mu} |A_\mu|^2 \right)^2 \left( e^{-S^{(2)}} - N^{-1} \right)^2.
\end{align}
\end{subequations}
\end{widetext}
In \cref{eqn:bound1}, we have defined $|A_\mu|$ to be the largest matrix element of $A_\mu$ in the Schmidt basis, and $\sqrt{p_\mathrm{min}}$ to be the smallest Schmidt coefficient. Taking a square root of both sides gives 
\begin{align}
    \Gmax \leq \frac{\sqrt{2} N}{p_\mathrm{min}} \left( \sum_{\mu} |A_\mu|^2 \right) \left( e^{-S^{(2)}} - N^{-1} \right),
\end{align}
recovering the result from the main text.

\subsection{Non-Full Rank Steady State}
If $\Psi$ is not full rank, then  $p_\mathrm{min}^{-1}$ is ill-defined. However, a similar bound can be derived for a state that is not full rank. Beginning at \cref{eqn:A2}, we note that
\begin{widetext}
\begin{align}
    \Gmax^2 &\leq \sum_{\mu,\nu} \left[  \sum_{j,k = 1}^N  \left( A_\mu)_{j,k} (A_\nu)_{j,k}^*   + (B_\mu)_{j,k} (B_\nu)_{j,k}^* \right) \left(  |\psi_j|^2 -  |\psi_k|^2 \right) \right]^2 \nonumber \\
    & \leq \sum_{\mu,\nu} \left( |A_\mu| |A_\nu|   + |B_\mu||B_\nu| \right)^2 \left[  \sum_{j,k = 1}^N \bigg|  |\psi_j|^2 -  |\psi_k|^2 \bigg| \right]^2 \nonumber \\
    &\leq N^2 \sum_{\mu,\nu} \left( |A_\mu| |A_\nu|   + |B_\mu||B_\nu| \right)^2 \sum_{j,k = 1}^N (  |\psi_j|^2 -  |\psi_k|^2 )^2 \nonumber \\
    &= 2N^3 \sum_{\mu,\nu} \left( |A_\mu| |A_\nu|   + |B_\mu||B_\nu| \right)^2 \left( e^{-S^{(2)}} - N^{-1} \right), \\
    \implies \Gmax &\leq 2N^{3/2} \sqrt{\sum_{\mu,\nu} \left( |A_\mu| |A_\nu|   + |B_\mu||B_\nu| \right)^2} \left( e^{-S^{(2)}} - N^{-1} \right)^{1/2} \nonumber \\
    & \leq  2 N^{3/2} \left[ \left(\sum_{\mu=1}^M |A_\mu| 
    \right)^2 + \left(\sum_{\mu=1}^M |B_\mu|  \right)^2
    \right] \delta \Erel^{1/2} ,
\end{align}
\end{widetext}
where we see that the bound depends only on the root of $\delta \Erel$ as opposed to linearly for the case of a full rank system.

\subsection{Necessity of Multiple Jump Operators}
\label{sec:Isospectral}
We now prove the result from the main text, that the condition \cref{eqn:A-B-Relation} implies that $A$ and $B$ are isospectral, necessitating multiple jump operators (or a Hamiltonian interaction) to get a unique steady state. Let's begin by assuming that the matrix $A$ is diagonalizable. The relation $A = -\Psi B^T \Psi^{-1}$ tells us that if $A$ is diagonalizeable via $A =P^{-1}DP$, then $B^T = -(P\Psi)^{-1} D (P \Psi)$. Hence, $B$ is also diagonalizeable as
\begin{align}
    B = (P\Psi)^{T} (-D) ((P \Psi)^{T})^{-1}.
\end{align}
Thus, we can work in the (non-orthonormal) basis of eigenvectors of $A$ and $B$ so that 
\begin{align}
    \hat L &= \hat D_A \otimes \1 - \1 \otimes \hat D_B.
\end{align}
and therefore every vector of the form $|\psi \rangle  = |i \rangle_A \otimes | i \rangle_B$ is in the kernel of $\hat L$, where $|i \rangle_{A,B}$ are the eigenbases of $A,B$, respectively. 

Alternatively, let's assume that $A$ is not diagonalizeable. In this case, we can find a basis where it is in Jordan-Normal form. Let's consider just a single Jordan Block of the form
\begin{align}
    A = \left(
    \begin{array}{cccccc}
       \lambda  & 1 & 0 & 0 & \dots & 0 \\
        0 & \lambda & 1 & 0  & \dots & 0\\
        \vdots & \vdots & \vdots & \ddots & \dots & 0 \\
        0 & 0 & 0 & \dots  & \lambda & 1 \\
        0 & 0 & 0 & \dots & 0 & \lambda         
    \end{array}
    \right)
\end{align}
of dimension $n \times n$. Now, we can rewrite this in the following way: $A$ has a single eigenvector we will denote as $|0 \rangle$ such that $A |0 \rangle = \lambda |0 \rangle$. Then, we define that $A|m \rangle = \lambda |m \rangle + |m - 1 \rangle$ for $m > 0$. Now, since $A, -B^T$ are similar matrices, they have an identical Jordan Normal form. Hence, we can define an identical basis for $B$ such that $B|m \rangle = -\lambda |m \rangle - |m - 1 \rangle$ (for $m > 0$) and $B|0 \rangle = \lambda |0 \rangle$. Now, consider the set of states 
\begin{align}
    |\phi_n \rangle &= \sum_{m = 0}^n |m \rangle \otimes |n - m \rangle.
\end{align}
Now, we can observe that $\hat L |\phi_n \rangle$ is
\begin{align}
    \hat L |\phi_n \rangle &= (A \otimes \1 + \1 \otimes B) \sum_{m = 0}^n |m \rangle \otimes |n - m \rangle \nonumber \\
    &= \sum_{m = 1}^n |m - 1 \rangle \otimes |n - m \rangle - \sum_{m = 0}^{n - 1}|m  \rangle \otimes |n - m - 1 \rangle \nonumber \\
    &= 0.
\end{align}
Repeating this construction for each Jordan block implies that the dimension of the kernel of $\hat L$ is always at least $N$. To lift this degeneracy of the Lindbladian, we need (at least) two jump operators such that $\ker \hat L_1 \cap \ker \hat L_2$ is spanned by the steady state $|\psi \rangle$, or a Hamiltonian interaction such that the intersection of the eigenvectors of the Hamiltonian and the kernel of $\hat L$ is spanned by the steady state $|\psi \rangle$.

In \cref{app:2} we will do this by choosing multiple random jump operators.

\subsection{Uneven Hilbert Space Dimension}

Throughout the main text, we mainly limit the discussion to bipartite Hilbert spaces where each subspace has an identical Hilbert space dimension. Now, we wish to explore what happens when considering systems of uneven dimension; without loss of generality we will assume $\mathcal{H} = \mathcal{H}_A \otimes \mathcal{H}_B$ and 
\begin{align}
    \dim(\mathcal{H}_A) \equiv N_A < N_B \equiv \dim(\mathcal{H}_B).
\end{align}
Now, we can still define a pure steady state in terms of it's Schmidt coefficients:
\begin{align}
    |\psi \rangle &= \sum_{i = 1}^{N_A} \sqrt{p_i} |i \rangle_A \otimes |i \rangle_B,
\end{align}
where $p_i$ are real and positive. We define the remaining $(N_B - N_A)$ dimensional subspace of $\mathcal{H}_B$ as
\begin{align}
    \mathcal{H}_B' &= \mathcal{H}_B \setminus \mathrm{span} \{|i \rangle_B | i = 1, \dots, N_A\},
\end{align}
for which we will define the basis $\{ |i\rangle_B | i = N_A + 1, \dots , N_B \}$. A maximally entangled state is still defined by a flat distribution of Schmidt coefficients, where $p_i = N_A^{-1}$. Taking $\hat L = \hat A \otimes \1 + \1 \otimes \hat B$ as before, we now calculate $|\hat L^\dagger | \psi \rangle|^2$ when the $\hat B$ is of larger rank than $\hat A$ and $|\psi \rangle$ is maximally entangled. We now find that
\begin{align}
    |\hat L^\dagger | \psi \rangle|^2 &= \langle \psi | [\hat L, \hat L^\dagger] | \psi \rangle \nonumber \\
    &= \langle \psi | [\hat A, \hat A^\dagger] \otimes \1 | \psi \rangle + \langle \psi | \1 \otimes [\hat B, \hat B^\dagger] | \psi \rangle \nonumber \\
    &= \tr [\hat A, \hat A^\dagger] + \frac{1}{N_A}\sum_{i = 1}^{N_A} \langle i| [\hat B, \hat B^\dagger] | i \rangle \nonumber \\
    &= \frac{1}{N_A}\sum_{i = 1}^{N_A} \sum_{j = N_A + 1}^{N_B} |B_{ij}|^2 - |B_{ji}|^2 \label{eqn:intermediate_result_1}.
\end{align}

\begin{figure}[t!]
    \centering
    \includegraphics[width = \linewidth]{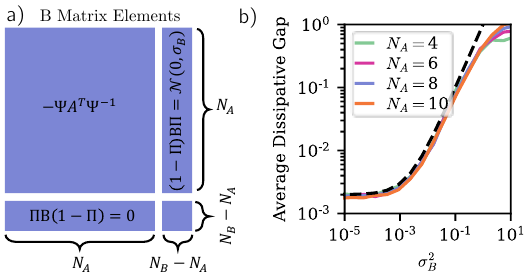}
    \caption{a) Depiction of the matrix elements of $B$ when there is an uneven Hilbert space dimension, breaking $B$ into four blocks using the projector $\Pi$ [c.f. \cref{eqn:B_projector}]. b) The dissipative gap for random Linbladians averaged over $100$ instances, with $\delta \Erel$ fixed to be $10^{-3}$. The $A$ matrix is sampled from the random Ginibre ensemble with unit variance, and the elements in $(\1 - \Pi)B\Pi$ are sampled from a normal distribution with zero mean and variance $\sigma_B^2$. the elements of $(\1 - \Pi)B(\1 - \Pi)$ are irrelevant, and taken to be 0. The black dashed line gives the expected scaling of $\sigma_B^2 + 2\delta \Erel$ until the gap saturates to an $\mathcal{O}(1)$ parameter for $\sigma_B^2 \gtrsim 1$.}
    \label{fig:6}
\end{figure}

\begin{figure*}[t!]
    \centering
    \includegraphics{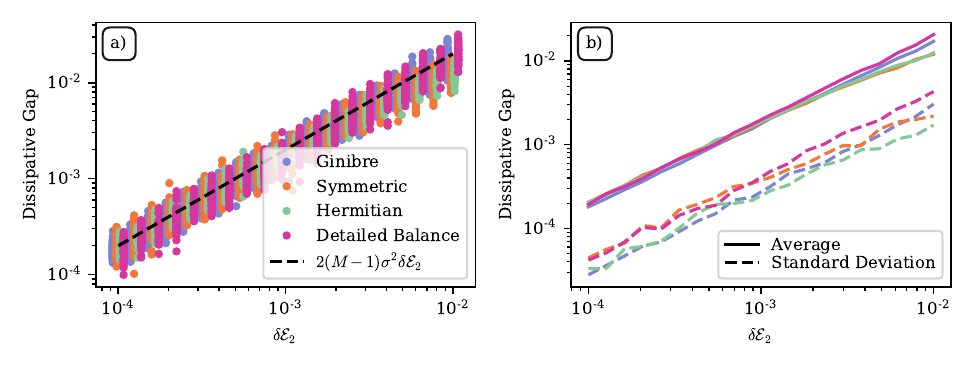}
    \caption{(a) Dissipative gap of random Lindbladians given a fixed steady state entanglement $\Erel$, similar to \cref{fig:2}. Different colors correspond to drawing the matrix $A$ from different distributions, see \cref{eqn:distributions}. For each distribution and entanglement value we sample 100 Lindbladians, with local dimension $N=10$ and $M=2$ jump operators. The black dashed line gives the scaling as predicted in \cref{eqn:HaarGap}. (b) Average (solid lines) and standard deviation (dashed lines) of points in (a) for a fixed value of entanglement and distribution.}
    \label{fig:5}
\end{figure*}

The condition $\hat L |\psi \rangle = 0$ implies (see \cref{fig:6})
\begin{align}
    B_{ij} &= \left\{
    \begin{array}{cc}
        A_{ji} \frac{\psi_i}{\psi_j} & i,j \leq N_A  \\
        0 & i > N_A, j \leq N_A
    \end{array}
    \right., \label{eqn:B_condition_uneven}
\end{align}
which allows us to reduce \cref{eqn:intermediate_result_1} to simply 
\begin{align}
     |\hat L^\dagger | \psi \rangle|^2 &= \frac{1}{N_A}\sum_{i = 1}^{N_A} \sum_{j = N_A + 1}^{N_B} |B_{ij}|^2,
\end{align}
which can now be non-zero; i.e. it is possible to generate a maximally entangled state if $N_B > N_A$, as was noted in the case of a qubit and qutrit in \cite{Brown2022}. Let's define the projection operator $\hat \Pi$ as
\begin{align}
    \hat \Pi &= \sum_{j= N_A + 1}^{N_B} |j \rangle \langle j|. \label{eqn:B_projector}
\end{align}

Then \cref{eqn:B_condition_uneven} tells us that $\hat \Pi \hat B (\1 - \hat \Pi) = 0$. However, we also know that to avoid slowdown, we require that $||(\1 - \hat \Pi) \hat B \hat \Pi || \neq 0$, as otherwise we can just truncate the Hilbert space and recover the bound for $N_A = N_B$. In fact, we can rewrite the bound in this case as
\begin{align}
    \Gmax \leq \frac{\sqrt{2} N}{p_\mathrm{min}} \left( \sum_{\mu} |A_\mu|^2 \right) \delta \Erel
    + \sum_\mu ||(\1 - \hat \Pi) \hat B_\mu \hat \Pi ||_2^2,
\end{align}
where the norm $|| \hat O ||_2$ is the operator norm defined as the largest singular value of $\hat O$. The first term is simply the standard one from before, and the second is a measure of how much the dissipation is utilizing the extra Hilbert space available. Using an uneven Hilbert space dimension to circumvent slowdown was first considered in a qubit-qutrit system in \cite{Brown2022}; however, it should also work perfectly well in the many body case. This can be seen in \cref{fig:6}, where we take $N_B = N_A + 1$, and take $A$ from the random Ginbre ensemble. Then we take the elements of $(\1 - \Pi)B\Pi$ to be normally distributed with zero mean and variance $\sigma_B$. Finally, we define the projection super-operator 
\begin{align}
    \hat{\mathcal{P}} &= \sum_{i,j = 1}^{N_A} |\rho_{ij} \rangle \rangle \langle \langle \rho_{ij} |, \\
    \hat \rho_{ij} &= |i \rangle \langle i |_A \otimes |j \rangle \langle j |_B.
\end{align}
(As before the double bracket notation $|\rho_{ij} \rangle \rangle$ signifies a vectorized density matrix.)
If $\L$ is the Lindbladian acting on the enlarged Hilbert space $(N_B > N_A)$, then we can define 
\begin{align}
    \L' &= \hat{\mathcal{P}} \L \hat{\mathcal{P}},
\end{align}
which gives the effective dynamics in the subspace of dimension $N_A^4$. If we look at the gap as a function of $\sigma_B^2$, we can see that when $\sigma_B \ll \delta \Erel$, then the gap is dominated by the entanglement induced slowdown. However, when $\sigma_B^2 \gtrsim  \delta S$, the gap opens up linearly in $\sigma_B^2$, before saturating at an $\mathcal{O}(1)$ value. Note the projection $\hat{\mathcal{P}}$ is necessary as otherwise when $\sigma_B^2 \ll 1$ the slow time-scale would be dominated by the time to get out of the extra $B$ system levels, and we would not see the plateau at small $\sigma_B^2$. 

An alternative interpretation is that systems on uneven Hilbert space dimension can be recast as effectively being of equal dimension with non-full-rank steady state Schmidt coefficients. That is to say, again assuming $N_B > N_A$, we can always define the operator 
\begin{align}
    \tilde A_\mu &= A_\mu \oplus 0_{N_B - N_A},
\end{align}
where by taking a direct sum with the $0$ matrix, we essentially just pad $A_\mu$ with zeroes so it is of the same dimension as $B_\mu$. Next, we define the steady state Schmidt coefficients
\begin{align}
    \tilde \psi_i &= \left\{ \begin{array}{cc}
       \psi_i   & i \leq N_A \\
        0 & N_A < i \leq N_B
    \end{array} \right\},
\end{align}
and again choose $B_\mu$ as in \cref{eqn:B_condition_uneven}. Now, we have a new master equation on an even Hilbert space dimension with identical dynamics. We have simply augmented the Hilbert space with uncoupled extra levels on the $A$ system. This means the steady state reduced density matrix is not full rank, but still the bound \cref{eq:ModBound} will apply. Hence, given a state on an uneven Hilbert space we can also state that 
\begin{widetext}
\begin{align}
    \Gmax \leq 2 N_B^{3/2} \left[ \left(\sum_{\mu=1}^M |A_\mu| 
    \right)^2 + \left(\sum_{\mu=1}^M |B_\mu|  \right)^2
    \right] \left( e^{-S^{(2)}_\mathrm{ss}} - \frac{1}{N_B} \right)^{1/2}.
\end{align}
\end{widetext}
In this case, $e^{-S^{(2)}_\mathrm{ss}} \geq N_A^{-1}$, and so the gap will not close even for a maximally entangled state, as shown in e.g. \cite{Brown2022}.

\subsection{Mixing Time}
Another relevant quantity in both classical or quantum markov chains is the mixing time \cite{temme2010} which can be thought of as the smallest time after which any initial probability distribution (or density matrix, in the quantum case) is within $\epsilon$ distance of the steady state distribution. More formally, we can define this as
\begin{align}
    t_{\mathrm{mix}}(\epsilon) &= \inf \{ t > 0 \ | \ \forall \hat \rho, d_{\mathrm{tr}}(e^{\L t} \hat \rho, \hat \rho_{\mathrm{ss}} ) \leq \epsilon \},
\end{align}
where $d_{\mathrm{tr}}$ is the trace distance and $\hat \rho_{\mathrm{ss}}$ is the steady state density matrix. The bound as stated in \cref{eqn:the_bound} is in terms of the quantum fidelity; however, this is related to the trace distance by \cite{nielsen2001}:
\begin{align}
    d_{\mathrm{tr}}(\hat \rho, \hat \sigma) \geq 1 - F(\hat \rho, \hat \sigma),
\end{align}
as long as at least one of $\hat \rho$ or $\hat \sigma$ is a pure state. Thus, if we now use that $F(e^{\L t} \hat \rho, \hat \rho_{\mathrm{ss}} ) \leq F (\hat \rho, \hat \rho_{\mathrm{ss}} ) + vt$, then we find that
\begin{align}
    \epsilon & \geq d_{\mathrm{tr}}(e^{\L t} \hat \rho, \hat \rho_{\mathrm{ss}} ) \geq 1 - F(e^{\L t} \hat \rho, \hat \rho_{\mathrm{ss}} ) \nonumber \\
    & \geq 1 - F (\hat \rho, \hat \rho_{\mathrm{ss}} ) - \Gmax t \\
    \implies & \Gmax t \geq 1 - \epsilon - F (\hat \rho, \hat \rho_{\mathrm{ss}} ) \\
    \implies & t_{\mathrm{mix}}(\epsilon) \geq \frac{1 - \epsilon}{\Gmax}
\end{align}
where $\Gmax$ is bounded from above by \cref{eqn:the_bound} as derived before. Hence, the mixing time is lower bounded by one over the distance from the maximal entropy.

\section{Random Lindbladians}
\label{app:2}

\subsection{Random Jump Operators}

It is useful to consider how well the bound is saturated on a class of random Lindbladians, as shown in the main text. Let's define a distribution of Schmidt coefficients $\{ p_i \}$ subject to
\begin{align}
    \sum_i p_i = 1, \ \ -\log \sum_i p_i^2 = S^{(2)},
\end{align}
for some fixed value of the Renyi-2 entropy $S^{(2)}$. Now, given this distribution, we define a set of random matrices $A_{\mu}$ sampled from the complex Ginibre ensemble where $(A_\mu)_{ij}$ are independent and identically distributed (i.i.d.) Gaussian random variables with
\begin{align}
    \mathds{E}[(A_\mu)_{ij}] &= 0, \ \ \mathds{E}[(A_\mu)_{ij}^* (A_\nu)_{kl}] = \delta_{ik} \delta_{jl} \delta_{\mu \nu} \sigma^2. \label{eqn:Ginibre}
\end{align}
We will sample random jump operators from many ensembles, including the complex Ginibre ensemble, random Hermitian matrices, random symmetric matrices, and random matrices which obey detailed balance. For concreteness, let's define $A$ drawn from the random Ginibre ensemble above [\cref{eqn:Ginibre}]. Then we define the random Hermitian operator $A_H$, random symmetric operator $A_S$, and random detailed balance operator $A_{DB}$ as 
\begin{subequations}
\label{eqn:distributions}
\begin{align}
    A_H &= \frac{1}{\sqrt{2}}(A  + A^\dagger), \\
    A_S &= \frac{1}{\sqrt{2}}(A + A^T), \\
    A_{DB} &= \frac{1}{\sqrt{2}}(A_S + \Psi A_S^T \Psi^{-1}). 
\end{align}
\end{subequations}
For all of these ensembles, we find that the dissipative gap scales linearly with  $\delta \Erel$ predicted by the bound. This is shown in \cref{fig:5}.

\subsection{Fidelity Rate of Change from a Haar Random State}

The bound \cref{eqn:the_bound} states that no state can approach the steady state at a rate faster than $\mathcal{O}(N^2) \times \delta \Erel$. However, numerics seem to imply that the dissipative gap is actually $\mathcal{O}(1) \times  \delta \Erel$. One might guess that this discrepancy is a result of the fact that dissipative gap is the slowest relaxing mode, whereas the bound applies to the fastest. To get a prefactor closer to $\mathcal{O}(1)$, we can consider bounding a Haar random state as opposed to all states. Recalling the formula from above [\cref{eqn:intermediate_result_2}], we know that
\begin{align}
    \partial_t F(t) &= \tr \left( \hat \rho_t \L^\dagger \rss \right) .
\end{align}
Let $\hat \rho_t = \hat U | 0 \rangle \langle 0 | \hat U^\dagger$ with $\hat U$ integrated over the Haar measure. Since $\partial_t F(t)$ is linear in $\hat \rho_t$, we can just directly calculate that
\begin{align}
    \int_{\mathrm{Haar}} \hat U|0 \rangle \langle 0| \hat U^\dagger \dd U &= \frac{\1}{N^2}; \\
    \implies \int_{\mathrm{Haar}}  \partial_t F(t) \dd U &= \frac{1}{N^2} \tr \left( \L^\dagger \rss \right).
\end{align}
Now, we can calculate that
\begin{align}
     &\frac{1}{N^2} \tr \left( \L^\dagger \rss \right) = \frac{1}{N^2} \sum_\mu \langle \psi | [\hat L_\mu, \hat L_\mu^\dagger] | \psi \rangle \nonumber \\
     &= \frac{1}{N^2} \sum_{\mu}   \sum_{j,k = 1}^N | ( A_\mu)_{j,k}|^2 \frac{\left(  p_j -  p_k \right)^2}{p_k}.  \label{eqn:intermediate_result_3}
\end{align}
Now at this point, we can see that this can be bounded in a similar way as before by
\begin{align}
    \int_{\mathrm{Haar}}  \partial_t F(t) \dd U \leq \left( \sum_\mu |A_\mu|^2 \right) \frac{2}{N p_\mathrm{min}} \delta \Erel. \label{eqn:intermediate_result_4}
\end{align}
We can be more precise if we know something about the distributions from which we sample $A_\mu$ and $p_j$. Suppose $(A_\mu)$ is sampled from a random distribution with variance
\begin{align}
    \mathds{E} \left[ | ( A_\mu)_{j,k}|^2 \right] &= \sigma^2.
\end{align}

Further, let's rewrite the Schmidt coefficients in the following, physically motivated form:
\begin{align}
    p_i &= \frac{e^{-\beta \lambda_i}}{\sum_{j = 1}^N e^{- \beta \lambda_j}}
\end{align}
where we have defined a new set of random variables $\lambda_i$. Note that by writing it in this manner, the fact that $\sum_{i = 1}^N p_i = 1$ is inherently manifest. Moreover, we have a single parameter $\beta$ that can tune the steady state entanglement entropy. When $\beta = 0$, then $p_i = 1/N$ and the state is maximally entangled. As $\beta \to \infty$, the state becomes pure. Moreover, $\delta \Erel(\beta)$ is monotonic in $\beta$, so there is a unique value of $\beta$ to fix the entanglement. Next, observe that if $\lambda_i \to \lambda_i - \overline{\lambda}$, then $p_i$ are invariant, so without loss of generality we can take $\sum_{i = 1}^N \mathds{E}[\lambda_i] = 0$. Next, we will set a scale for $\beta$ by defining
\begin{align}
    \frac{1}{N} \sum_{i = 1}^N \mathds{E}[\lambda_i^2] &= 1
\end{align}
We can always do this (for any distribution with a well defined second moment) by rescaling $\beta$. Finally, we will define the variable $\chi^2 = \sum_{i,j=1}^N \mathds{E}[\lambda_i \lambda_j]$. For example, if $\lambda_i$ are i.i.d. then $\chi^2 = N$. To calculate the average of the rate of change of the fidelity, it will be necessary to compute the average $\sum_{i = 1}^N \mathds{E}[p_i^{-1}]$. We can make progress by noting that, in the high entanglement limit, the variance of the $p_i$ is highly constrained by fixing the entropy. Hence, we can compute this to leading order in $\delta \Erel$ (or equivalently, the small $\beta$ limit). Observe that
\begin{align}
    \mathds{E}[\delta \Erel] &= \mathds{E} \left[ \frac{\sum_{i = 1}^N e^{-2\beta \lambda_i}}{ \left( \sum_{j = 1}^N e^{- \beta \lambda_j}\right)^2 } \right] \nonumber \\
    &= \frac{\beta^2}{N} \left( 1 - \frac{\chi^2}{N^2} \right) + \mathcal{O}(\beta^4 N^2) 
\end{align}
Now, $\delta \Erel$ is in fact fixed, so we can invert this to be an equation instead for $\beta$:
\begin{align}
    \frac{\beta^2}{N} \left( 1 - \frac{\chi^2}{N^2} \right) &= \delta \Erel + \mathcal{O} (\delta \Erel N^2)^2
\end{align}
From here, we can now calculate, to leading order in $\beta$:
\begin{align}
\sum_{i=1}^N \mathds{E}[p_i^{-1}] &= \sum_{i,j = 1}^N \mathds{E} \left[  e^{-\beta(\lambda_i - \lambda_j)} \right] \nonumber \\
&= N^2 + \frac{\beta^2}{2} \sum_{j,k = 1} \mathds{E} \left[  (\lambda_i - \lambda_j)^2 \right] + \mathcal{O}(\beta^4 N^2) \nonumber \\
&= N^2 \left( 1 + N \delta \Erel \right) + \mathcal{O}(N^2 \delta \Erel)^2
\end{align}

We can use this relation to observe that \cref{eqn:intermediate_result_3} simplifies to
\begin{align}
    &\mathds{E} \left[ \int_{\mathrm{Haar}}  \partial_t F(t) \dd U \right] \nonumber \\
    &= \mathds{E} \left[ \frac{1}{N^2} \sum_{\mu = 1}^M   \sum_{j,k = 1}^N | ( A_\mu)_{j,k}|^2 \frac{\left(  p_j - p_k \right)^2}{p_k}   \right] \nonumber \\
    &= \frac{M \sigma^2}{N^2} \mathds{E} \left[   \sum_{j,k = 1}^N \frac{\left(  p_j - p_k \right)^2}{p_k}    \right] \nonumber \\
    &= 2M \sigma^2 \delta \Erel + \mathcal{O}(\delta \Erel)^2,
\end{align}
where we defined $M$ to be the number of jump operators. Assuming $N \delta \Erel \ll 1$, we can drop the second term as small and recover the result quoted in the text. We can also calculate the variance one would expect from such a distribution. Here, we calculate 
\begin{align}
    &\mathds{E} \left[ \int_{\mathrm{Haar}}  |\partial_t F(t)|^2 \dd U \right] \nonumber \\
    &= \mathds{E} \left[ \int_{\mathrm{Haar}} \left| \sum_\mu |\langle \psi | L_\mu U |0 \rangle|^2  \right|^2 \dd U \right].
\end{align}
This now depends nonlinearly on $\hat \rho_t$, so we cannot simply replace the state with its average. However, we can still make progress. We will suppress the integral over the Haar measure for brevity, giving
\begin{widetext}
\begin{align}
   \mathds{E} \left[ |\partial_t F|^2 \right] &= \mathds{E} \left[ \left| \sum_\mu |\langle \psi | \hat L_\mu \hat U |0 \rangle|^2  \right|^2 \right] \nonumber \\
   &= \mathds{E} \left[ \sum_{i_\alpha, j_\alpha} ^N \sum_{\mu, \nu}^m (A_{\mu})_{j_1k_1} (A_{\nu})_{j_2k_2} (A_{\mu})_{j_3k_3}^* (A_{\nu})_{j_4k_4}^*  U^{k_1j_1}_{00}U^{k_2j_2}_{00} (U^{k_3j_3}_{00})^* (U^{k_4j_4}_{00})^* \prod_{\alpha} \left( \sqrt{p_{j_\alpha}} -  \frac{p_{k_\alpha}}{\sqrt{p_{j_\alpha}}} \right) \right] \nonumber \\
    &= \sigma^4 \sum_{i_\alpha, j_\alpha} ^N \sum_{\mu, \nu}^m \left( \delta_{13} \delta_{24} + \delta_{14} \delta_{23} \delta_{\mu \nu} \right) \mathds{E} \left[   U^{k_1j_1}_{00}U^{k_2j_2}_{00} (U^{k_3j_3}_{00})^* (U^{k_4j_4}_{00})^* \prod_{\alpha} \left( \sqrt{p_{j_\alpha}} -  \frac{p_{k_\alpha}}{\sqrt{p_{j_\alpha}}} \right) \right] \nonumber \\
    &= \sigma^4 \sum_{i_\alpha, j_\alpha}^N \left(M^2 \delta_{13} \delta_{24} + m \delta_{14} \delta_{23} \right) \frac{\left(\delta_{13} \delta_{24} + \delta_{14} \delta_{23} \right)}{N^2(N^2 + 1)}  \mathds{E} \left[ \prod_{\alpha} \left( \sqrt{p_{j_\alpha}} -  \frac{p_{k_\alpha}}{\sqrt{p_{j_\alpha}}} \right) \right] \nonumber \\
    &= \frac{\sigma^4(M^2 + M)}{N^2(N^2 + 1)} \sum_{i_\alpha, j_\alpha}^N \left( \delta_{13} \delta_{24} + \delta_{1234} \right) \mathds{E} \left[ \prod_{\alpha} \left( \sqrt{p_{j_\alpha}} -  \frac{p_{k_\alpha}}{\sqrt{p_{j_\alpha}}} \right) \right] \nonumber \\
    &= \frac{\sigma^4(M^2 + M)}{N^2(N^2 + 1)} \mathds{E} \left[ \left( \sum_{jk}^N \frac{(p_j - p_k)^2}{p_j} \right)^2 + \sum_{jk}^N \frac{(p_j - p_k)^4}{p_j^2} \right].
\end{align}
\end{widetext}
To make progress, we will again assume that $\delta \Erel \ll N^{-2}$, and so we can again expand to leading order in $\delta \Erel N^2$. We will also assume $N \gg 1$, and so we will only keep terms to leading order in $N^{-1}$, as well.
\begin{figure}[t]
    \centering
    \includegraphics[width = \linewidth]{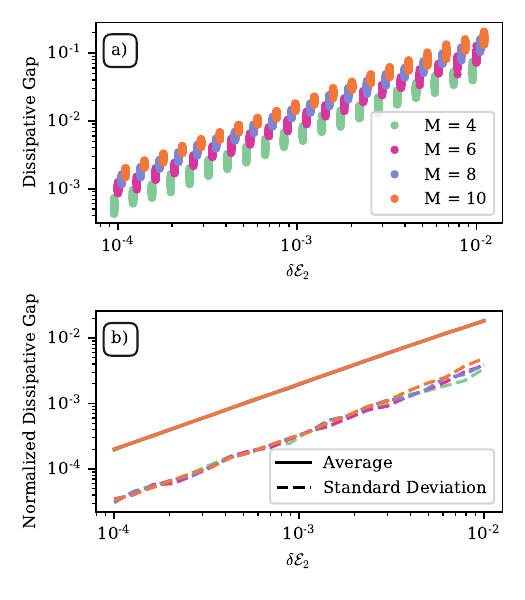}
    \caption{
    (a) Dissipative gap of random Lindbladians with $A_\mu$ sampled from the complex Ginibre ensemble and local dimension $N = 10$ for varying numbers of jump operators $M$. There are 100 samples for each $M$ and each value of entanglement $\delta \Erel$.
    (b) The normalized average (solid lines) and standard deviation (dashed lines) of data in (a). The average is normalized as $\mathds{E}[\Delta]/(m-1)$ [c.f. \cref{eqn:mean_gap}], and the standard deviation as $\sqrt{ \frac{ \mathds{E}[\Delta^2] - \mathds{E}[\Delta]^2 }{m-1} }$ [c.f. \cref{eqn:var_gap}].
    }
    \label{fig:7}
\end{figure}
This gives
\begin{align}
    \mathds{E} \left[ |\partial_t F|^2 \right] &= \sigma^4(M^2 + M + \mathcal{O}(N^{-2})) \left( \delta \Erel \right)^2 + \mathcal{O}(N^2 \delta \Erel)^3. \label{eqn:var_scaling}
\end{align}
This tells us that we would expect a variance of 
\begin{align}
    \mathds{V} \left[ |\partial_t F| \right] &= \mathds{E} \left[ |\partial_t F|^2 \right] - \left| \mathds{E} \left[ |\partial_t F| \right] \right|^2 \nonumber \\
    &= \sigma^4 M \left( \delta \Erel \right)^2   + \mathcal{O}(\delta \Erel)^3 \nonumber \\
    &= \frac{1}{M} \left| \mathds{E} \left[ |\partial_t F| \right] \right|^2  +   \mathcal{O}(\delta \Erel)^3,
\end{align}
and so the distribution should get tighter as one adds more and more jump operators. This scaling can be observed in \cref{fig:7}. However, we note that in \cref{fig:7} we are plotting the dissipative gap and not the rate of change of the fidelity. This is important because we know analytically that the gap $\Delta = 0$ if $M=1$, given the fact that $\hat A$ and $\hat B$ are isospectral (see \cref{sec:Isospectral}). However, a single jump operator is sufficient to change the fidelity to the steady state at a non-zero rate for some states in the Hilbert space. Hence, we expect that the average and variance of the gap should be (setting $\sigma = 1$):
\begin{align}
    \mathds{E}[\Delta] &\propto (M-1) \delta \Erel, \label{eqn:mean_gap} \\
    \mathds{V}[\Delta] &\propto  (M-1) \left( \delta \Erel \right)^2. \label{eqn:var_gap}
\end{align}
Hence, in \cref{fig:7} we normalize by these values and observe a collapse of both the average and the standard deviation (root of the variance) onto a single line. 

\section{Maximally Entangled State as a Strong Symmetry}
\label{app:3}

Recall from \cref{app:1} that if $\rss$ is a maximally entangled state, then $\rss \hat L_\mu = 0$. However, because to be a steady state requires $\hat L_\mu \rss = 0$, this implies that the commutator of the steady state with each jump operator identically vanishes. Additionally, any pure steady state must be a Hamiltonian eigenstate as well, so all together this implies:
\begin{align}
    [\hat L_\mu, \rss] = [\hat H, \rss] = 0.
\end{align}
However, this is simply the statement that $\rss$ is a \textit{strong symmetry} of the dynamics \cite{Buca2012}. This means that we can interpret the steady state density matrix as a symmetry operator; i.e., it is a conserved charge. Because it is a pure state $\rss = |\psi \rangle \langle \psi |$, the density matrix is itself simply a projection operator onto the steady state $|\psi \rangle$, and so the conserved charge generated by such a symmetry is simply the population in the steady state. Hence, this is another way to see that if an open system has a steady state that is maximally entangled, then it is completely dynamically isolated from the rest of the system.

Now, let's suppose we perturb the Lindbladian slightly away from this point, so that the steady state is still pure, but slightly less entangled. In this case we still have that
\begin{align}
    \hat L_\mu \rss = [\hat H, \rss] = 0,
\end{align}
since it is by definition a pure steady state, but if it is not maximally entangled then generically $\rss \hat L_\mu \neq 0$. We can quantify, then, how close this is to being a symmetry by defining the error $\mathcal{E}$
\begin{align}
    \mathcal{E} &= \sum_\mu |[\hat L_\mu, \rss]| = \sum_\mu \langle \psi | [\hat L_\mu, \hat L_\mu^\dagger ] | \psi \rangle = \tr \left( \L^\dagger \rss \right), \label{eqn:symError}
\end{align}
where $|\cdot |$ denotes the Hilbert-Schmidt norm. However, this is exactly the same term that shows up when calculating the rate of change of the fidelity, which we know goes to zero as [c.f. \cref{eqn:intermediate_result_3,eqn:intermediate_result_4}]
\begin{align}
     \sum_\mu \langle \psi | [\hat L_\mu, \hat L_\mu^\dagger ] | \psi \rangle  \lesssim N^2 \delta \Erel,
\end{align}
and so we can think of this entanglement term both as bounding how fast the fidelity to the steady state can change, as well as how close the Lindbladian is to having a strong symmetry.

This interpretation also explains why there is exactly one midgap state for a random Lindbladian. Let's suppose that we have a Lindbladian $\L_0$ with a maximally entangled steady state $\hat \rho_0 = |\psi_{\mathrm{max}} \rangle \langle \psi_{\mathrm{max}}|$. Now, we know that this implies that there is a strong symmetry in the dynamics, and therefore at least two-fold degeneracy in the steady state manifold. Moreover, in the absence of any other symmetry constraints, the degeneracy should be exactly two, and it should be split by a $\Delta_{\mathrm{bulk}} = \mathcal{O}(1)$ bulk gap \cite{Denisov2019}. Now, let's assume that there is a very closely related Lindbladian $\L = \L_0 + \epsilon \L_1$ such that $\epsilon \ll \Delta_{\mathrm{bulk}}$ the bulk gap, and so we can perform perturbation theory within the steady state manifold. 

Now, we don't know exactly what the degenerate steady state actually is, but we do know exactly what the \textit{left eigenvectors} are, so instead we will do perturbation theory in $\L^\dagger$, which has the same eigenspectrum of $\L$. Explicitly, we know that $\L_0^\dagger \hat \rho_0 = \L_0^\dagger (\1 - \hat \rho_0) = 0$. 

Let's suppose that $\rss$ is the unique steady state solution of $\L \rss = 0$. Then $\rss = \hat \rho_0 + \epsilon \hat \rho_1 + \mathcal{O}(\epsilon^2)$. Therefore, $\hat \rho_0 = \rss + \mathcal{O}(\epsilon)$, and specifically this means that the left eigenvector of $\L$, or alternatively the right eigenvector of $\L^\dagger$ is $\hat \rho_0 + \mathcal{O}(\epsilon) = \rss + \mathcal{O}(\epsilon)$. Hence, to first order in $\epsilon$, we can perturbatively calculate the steady state degeneracy splitting as the eigenvalues of
\begin{align}
    &\tr \left[ \left( \begin{array}{c}
       \1 - \rss \\
       \rss 
    \end{array}
    \right)
    \L^\dagger 
    \left( 
    \begin{array}{cc}
       \1 - \rss  & \rss 
    \end{array}
    \right)
    \right] \nonumber  \\
    &=     \tr \left[ \left( \begin{array}{c}
       \1  \\
       0 
    \end{array}
    \right)
    \L^\dagger 
    \left( 
    \begin{array}{cc}
       - \rss  & \rss 
    \end{array}
    \right)
    \right] = \left( 
    \begin{array}{cc}
       - \Gamma & \Gamma \\
       0 & 0 
    \end{array}
    \right), \label{eqn:dissGapOrder}
\end{align}
where $\Gamma = \tr (\L^\dagger \rss)$. The eigenvalues of this matrix are $0,-\Gamma$, so perturbatively one would expect a dissipative gap that is equivalent to $ \Gamma = \tr( \L^\dagger \rss )$. However, this is just exactly the error $\mathcal{E}$ defined in \cref{eqn:symError}; that is to say, the dissipative gap is equivalent to how close the system is to having a strong symmetry.

\section{Directional Dynamics}
\label{app:4}

As mentioned in the text, the form of the jump operator \cref{eqn:me3} is the correct form for directional dynamics \cite{Metelmann2015,Carmichael1993,Gardiner1993,Stannigel2012}. To achieve directionality, such that the dynamics in the $A$ system are unaffected by those in the $B$ system, a Hamiltonian interaction is necessary. Generically, given a jump operator $\hat L = \hat A \otimes \1 + \1 \otimes \hat B$, one can always make this a unidirectional (chiral) process via the Hamiltonian \cite{Metelmann2015,Stannigel2012}
\begin{align}
    \hat H_{AB} &= \frac{i}{2} \left( \hat A^\dagger \otimes \hat B - \hat A \otimes \hat B^\dagger \right).
\end{align}
More is needed, however, to ensure that the state we began with is still the steady state of this new Lindbladian.

To make this possible, firstly, we will need a local Hamiltonian $\hat H_A$ on the $A$ system such that
\begin{align}
    \L_A \hat \rho_A &= -i[\hat H_A, \hat \rho_A] + \D[\hat A] \hat \rho_A = 0, \\
    \hat \rho_A & \equiv  \tr_B \rss .
\end{align}
This is equivalent to the condition that 
\begin{align}
    (H_A)_{nm} &= \frac{i}{p_m - p_n} \langle n | \D[\hat A] \hat \rho_A | m \rangle,
\end{align}
which uniquely defines the matrix elements of $H_A$ in the Schmidt basis. Note that this also gives another constraint, that 
\begin{align}
    \langle n | \D[\hat A] \hat \rho_A | n \rangle &= 0 ,\\
    \implies \sum_k |A_{nk}|^2 p_k - |A_{kn}|^2 p_n &= 0.
\end{align}
One way to satisfy this is to simply assume $A = \Psi A^T \Psi^{-1}$, which is what we have done throughout this manuscript. 

With $\hat H_A$ now defined, we can use the general CQA construction in \cite{Stannigel2012} to find the local Hamiltonian on $B$, which is given by
\begin{align}
    H_B &= -\frac{1}{2} \Psi \left( H_A - \frac{i}{2} A^\dagger A \right)^T \Psi^{-1} + \hc
\end{align}
Then, we can define 
\begin{align}
    \hat H_{\mathrm{CQA}} &= \hat H_A \otimes \1 + \1 \otimes \hat H_B + \hat H_{AB}, \label{eqn:CQA_Ham} \\
    \mathcolorbox{white}{\L_{\mathrm{CQA}} \hat \rho} &\mathcolorbox{white}{= \D[\hat A \otimes \1 + \1 \otimes \hat B]\hat \rho  -i[\hat H_{\mathrm{CQA}},\hat \rho],}
\end{align}
which generates directional dynamics with the steady state $\rss$ as desired.

\section{Free-Fermion Subspaces of an Interacting Hamiltonian}
\label{app:5}

In the main text, we consider the interacting, two-leg ladder Hamiltonian $\hat H = \hat H_{\parallel} + \hat H_{\perp}$, where $\hat H_\parallel$ is an XX Hamiltonian running along the legs of the ladder, and $\hat H_\perp$ is an anisotropic XXZ Hamiltonian across the rungs:
\begin{subequations}
\begin{align}
    \hat H_{\parallel} &= J\sum_{i = 1}^{n-1} \sum_{s = A,B} \hat \sigma_{s,i}^+ \hat \sigma_{s, i + 1}^- + \hc, \\
    \hat H_{\perp} &= \sum_{i = 1}^{n} J \left( \hat \sigma^x_{A,i} \hat \sigma^x_{B,i} + \hat \sigma^y_{A,i} \hat \sigma^y_{B,i} \right) + J_z \hat \sigma^z_{A,i} \hat \sigma^z_{B,i}, 
\end{align}
\end{subequations}
This Hamiltonian was introduced in Ref. \cite{Znidaric2013}, where it was shown that (see also \cite{lingenfelter2023}) the Hamiltonian can be recast as particles hopping on a 1D chain, where now each lattice site has local Hilbert space dimension $d = 4$. Define the singlet and triplet states $|S/T \rangle_j$ as
\begin{align}
    |S/T \rangle_j &= \frac{1}{\sqrt{2}} \left( |0\rangle_{A,j} |1 \rangle_{B,j} \mp |1\rangle_{A,j} |0 \rangle_{B,j} \right).
\end{align}
Letting $|0\rangle_j \equiv |0\rangle_{A,j} |0\rangle_{B,j}$ and similarly $|1\rangle_j \equiv |1\rangle_{A,j} |1\rangle_{B,j}$, then $\{ |0\rangle, |1 \rangle, |S \rangle, |T \rangle \}$ span the local Hilbert space of the single 1D chain. 

Next, it is simple to observe that because the XXZ interaction conserves total angular momentum as well as total Z angular momentum on each bond, then these four states can also be used to form an eigenbasis of $\hat H_\perp$. It is also of interest to note that the states
\begin{align}
    |\Omega_1 \rangle & \equiv |S\rangle_1 \otimes |T\rangle_2 \otimes \dots \otimes  |S(T) \rangle_n, \\
    |\Omega_2 \rangle & \equiv |T\rangle_1 \otimes |S\rangle_2 \otimes \dots \otimes |T(S) \rangle_n,
\end{align}
are also zero energy eigenstates of $\hat H_\parallel$, and so we will define these to be the ``vacua'' of the Hamiltonian $\hat H$.

From here, one can add ``excitations'' in the form of $|0\rangle$ or $|1\rangle$. This is because the Hamiltonian $\hat H_\parallel$ sends (c.f. Eq. 2 in \cite{Znidaric2013})

\begin{subequations}
    \begin{align}
    |0S\rangle \leftrightarrow |S0 \rangle, \ \ \ \ |0T\rangle \leftrightarrow |T0 \rangle, \\
    |1S\rangle \leftrightarrow |S1 \rangle, \ \ \ \ |1T\rangle \leftrightarrow |T1 \rangle, 
\end{align}
\end{subequations}
where $\leftrightarrow$ represents mapping under $\hat H_\parallel$ modulo multiplicative constants. 
Furthermore, $\hat H_\parallel |00 \rangle = \hat H_\parallel |11 \rangle = 0$, and so any fixed number of $|0\rangle$ \textit{or} $|1\rangle$ particles on top of the vacuum can be mapped exactly to free fermions - each individual species experiences a nearest neighbor hopping Hamiltonian without scattering. The caveat here, and the reason that the entire Hilbert space is not equivalent to free fermions, is that the $|0\rangle$ particles and $|1\rangle$ particles can scatter off of each other. I.e. $\hat H_\parallel$ maps (c.f. Eq. 3 in \cite{Znidaric2013})
\begin{align}
    |01\rangle + |10 \rangle \leftrightarrow |TT \rangle - |SS \rangle.
\end{align}
Hence, we can observe that for a given ground state and a given particle species, there are $\sum_{i = 0}^n \binom{n}{i} = 2^n$ different states in the free-fermion subspace, giving overall a $2^{n + 2} - 4$ dimensional Hilbert space. On the other hand, the total Hilbert space dimension is $2^{2n}$, so the free-fermion sector is exponentially large in $n$, but still exponentially small compared to the full space.

We now demonstrate that the state 
\begin{align}
    |\psi \rangle_\mathrm{ss} = \bigotimes_{i = 1}^n \left[\sqrt{1-v^2}|0\rangle_{A,i} |0\rangle_{B,i} + (-1)^i v|1\rangle_{A,i} |1 \rangle_{B,i} \right]. \label{eqn:rainbow_steadystate_2}
\end{align}
is an eigenstate of $\hat H_\parallel + \hat H_\perp$, and therefore a steady state of the Liouvillian 
\begin{align}
    \mathcolorbox{white}{\L \hat \rho} & \mathcolorbox{white}{= -i[\hat H_\parallel + \hat H_\perp, \hat \rho] + \L_\mathrm{diss} \hat \rho }
\end{align}
with $\L_\mathrm{diss}$ as given in \cref{eqn:local_jump} in the main text. We have already noted that $\hat H_\parallel |\psi \rangle_\mathrm{ss} = 0$, so it remains only to show that $\hat H_\perp |\psi \rangle_\mathrm{ss} = nJ_z|\psi \rangle_\mathrm{ss}$. By direct computation, we can observe that
\begin{widetext}
\begin{align}
     \hat H_{\perp}|\psi \rangle_\mathrm{ss} &= \left[ \sum_{i = 1}^{n} J \left( \hat \sigma^x_{A,i} \hat \sigma^x_{B,i} + \hat \sigma^y_{A,i} \hat \sigma^y_{B,i} \right) + J_z \hat \sigma^z_{A,i} \hat \sigma^z_{B,i}\right]  \bigotimes_{j = 1}^n \left[\sqrt{1-v^2}|0\rangle_{A,j} |0\rangle_{B,j} + (-1)^j v|1\rangle_{A,j} |1 \rangle_{B,j} \right] \nonumber \\
     &= \left[ \sum_{i = 1}^{n}  J_z \hat \sigma^z_{A,i} \hat \sigma^z_{B,i}\right]  \bigotimes_{j = 1}^n \left[\sqrt{1-v^2}|0\rangle_{A,j} |0\rangle_{B,j} + (-1)^j v|1\rangle_{A,j} |1 \rangle_{B,j} \right] \nonumber \\
     &= \left[ \sum_{i = 1}^{n} J_z \right]  \bigotimes_{j = 1}^n \left[\sqrt{1-v^2}|0\rangle_{A,j} |0\rangle_{B,j} + (-1)^j v|1\rangle_{A,j} |1 \rangle_{B,j} \right] =  nJ_z|\psi \rangle_\mathrm{ss}
\end{align}
\end{widetext}
as desired.

\section{Derivation of Local Lindblad Master Equations from System-Bath Coupling }
\label{app:6}
Here, we will derive how one gets a Lindblad-style master equation with jump operators respecting the locality constraint [c.f.~\cref{eqn:me3}] starting from a system-bath coupling of the form in \cref{eqn:sys-bath-coupling}in the main text. We will follow the general procedure of taking first the Born-Markov and then Rotating Wave approximations, as laid out in, e.g., Refs.~\cite{Breuer2002,Gardiner2004}. We will not thoroughly justify each approximation, as this is explained in great detail already elsewhere; for these details we encourage the reader to consult Refs.~\cite{Breuer2002,Gardiner2004}. We begin by assuming that we have a tripartite quantum system, with a Hilbert space $\mathcal{H} = \mathcal{H}_R \otimes \mathcal{H}_A \otimes \mathcal{H}_B$. Overall, these describe a Reservoir ($\mathcal{H}_R$) which will be traced out to give the open system dynamics, along with a bipartite system composed of $\mathcal{H}_{A,B}$. The full Hamiltonian can be written in the form
\begin{align}
    \mathcolorbox{white}{\hat H} &\mathcolorbox{white}{= \hat H_R + \hat H_S + \hat H_I, }
\end{align}
with $\hat H_R$ giving local dynamics of the reservoir, $\hat H_S$ local dynamics of the bipartite system, and $\hat H_I$ giving their (weak) interaction. We will assume that the local dynamics $\hat H_S$ can be further decomposed $\hat H_S = \hat H_A + \hat H_B$, so that there is no explicit interaction between subsytems not mediated by the reservoir. 

Now, let's define $\hat{\tilde{\chi}}(t)$ to be the density matrix for the entire system plus reservoir. It is governed by an equation of motion
\begin{align}
    \mathcolorbox{white}{\partial_t\hat{\tilde{\chi}} }&\mathcolorbox{white}{= -i[\hat H, \hat{\tilde{\chi}}]. }
\end{align}
If we work in the interaction picture of the local dynamics, we can define $\U = \exp\left( it \left(\hat H_R + \hat H_S \right) \right)$ so that $\hat \chi(t) = \U \hat{\tilde{\chi}}(t) \U^\dagger $ obeys simply
\begin{align}
    \mathcolorbox{white}{\partial_t \hat \chi} &\mathcolorbox{white}{= -i[\hat H_I(t), \hat \chi]. }
\end{align}
From here, we can formally integrate the equation of motion to find that
\begin{align}
    \mathcolorbox{white}{\hat \chi(t) }&\mathcolorbox{white}{= \hat \chi(0) - i \int_0^t [\hat H_I(s), \hat \chi(s)] \dd s.}
\end{align}
At this point, we can define the system density matrix $\hat \rho$ as the partial trace over the reservoir degrees of freedom: $\hat \rho = \tr_R \hat \chi$. This gives an equation of motion
\begin{align}
    \mathcolorbox{white}{\partial_t \hat \rho} &\mathcolorbox{white}{= - \int_0^t \tr_B [\hat H_I(t), [\hat H_I(s), \hat \chi(s)]] \dd s. }
\end{align}
To this point, the equation is exact. However, to make progress, we will make the Born-Markov approximation that the full density matrix $\hat \chi(t) \approx \hat \rho(t) \otimes \hat \rho_R$ for all times: i.e. the reservoir is always approximately in the same state $\rho_R$, which we take to be static (diagonal in the eigenvectors of $\hat H_R$, for example a thermal state). Next, we will assume that the local time dynamics only depend on the state at the given time, i.e. there is no memory effect. This allows us to replace $\hat \chi(s) \to \hat \chi(t)$ in the integral, giving the new equation of motion
\begin{align}
    \mathcolorbox{white}{\partial_t \hat \rho }&\mathcolorbox{white}{= - \int_0^t \tr_B [\hat H_I(t), [\hat H_I(t-s), \hat \rho(t) \otimes \hat \rho_R]] \dd s.}
\end{align}
The fact that we replaced $\hat \chi(t)$ with $\hat \chi(s)$ tells us that the integration kernel should be tightly peaked around $|t-s| \sim 0$, and so we can extend the upper integration bound to infinity with very little error. This gives finally the time-local Redfield equation:
\begin{align}
    \mathcolorbox{white}{\partial_t \hat \rho }&\mathcolorbox{white}{= - \int_0^\infty \tr_B [\hat H_I(t), [\hat H_I(t-s), \hat \rho(t) \otimes \hat \rho_R]] \dd s. }\label{eqn:BornMarkov}
\end{align}
To get something in Lindblad form, we must now make the rotating wave approximation. For this it will be necessary to recall the exact form of the interaction Hamiltonian:
\begin{align}
    \mathcolorbox{white}{\hat H_I} &\mathcolorbox{white}{= \sum_{\mu=1}^M \hat R_{A,\mu} \otimes \hat{\tilde{A}}_\mu \otimes \1 + \hat R_{B,\mu} \otimes \1 \otimes \hat{\tilde{B}}_\mu + \hc }\nonumber \\
    &\mathcolorbox{white}{= \sum_{j = 1}^{4M} \hat R'_j \otimes \hat S'_j, }
\end{align} 
\begin{align}
    \mathcolorbox{white}{\hat R'_j} &\mathcolorbox{white}{= \left\{ 
    \begin{array}{cc}
       \frac{1}{\sqrt{2}}(\hat R_{A,j} + \hat R_{A,j}^\dagger)  &  1 \leq j \leq M\\
       \frac{i}{\sqrt{2}}(\hat R_{A,j-M} - \hat R_{A,j-M}^\dagger)  &  M < j \leq 2M\\
       \frac{1}{\sqrt{2}}(\hat R_{B,j - 2M} + \hat R_{B,j-2M}^\dagger)  &  2M < j \leq 3M\\
       \frac{i}{\sqrt{2}}(\hat R_{B,j-3M} - \hat R_{B,j-3M}^\dagger)  &  3M < j \leq 4M\\
    \end{array}
    \right.}
\end{align} 
\begin{align}
    \mathcolorbox{white}{\hat S'_j }&\mathcolorbox{white}{= \left\{ 
    \begin{array}{cc}
       \frac{1}{\sqrt{2}}(\hat{\tilde{A}}_j + \hat{\tilde{A}}_j^\dagger) \otimes \1  &  1 \leq j \leq M\\
       \frac{i}{\sqrt{2}}(\hat{\tilde{A}}_{j-M} - \hat{\tilde{A}}_{j-M}^\dagger) \otimes \1  &  M < j \leq 2M\\
       \frac{1}{\sqrt{2}} \1 \otimes (\hat{\tilde{B}}_{j-2M} + \hat{\tilde{A}}_{j-2M}^\dagger)   &  2M < j \leq 3M\\
       \frac{i}{\sqrt{2}} \1s \otimes (\hat{\tilde{B}}_{j-3M} - \hat{\tilde{A}}_{j-3M}^\dagger)   &  3M < j \leq4M\\
    \end{array}
    \right. }\label{eqn:localSOPs}
\end{align}
where we have defined $\hat R'_j, \hat S'_j$ to be Hermitian operators that act on either the reservoir or the system, respectively. It will also be extremely important to observe that each system operator is local to either the $A$ or $B$ subsystem. Next, we will decompose the system operators $\hat S_j'$ into their frequency components. If we define the operator $\hat \Pi_\epsilon$ as the projector onto the eigenspace of $\hat H_S$ with eigenvalue $\epsilon$, then we can rewrite
\begin{align}
    \mathcolorbox{white}{\hat S_j' }&\mathcolorbox{white}{= \sum_\omega \hat S_j'(\omega),} \\
    \mathcolorbox{white}{\hat S_j'(\omega) }&\mathcolorbox{white}{= \sum_{\epsilon - \epsilon' = \omega} \hat \Pi_{\epsilon'} \hat S_j' \hat \Pi_\epsilon.}
\end{align}
Using these operators, we can now expand the equation of motion as
\begin{widetext}
    \begin{align}
    \mathcolorbox{white}{\partial_t \hat \rho} &\mathcolorbox{white}{= \sum_{\omega, \omega', j,k} e^{i(\omega - \omega')t} C_{jk}(\omega) \left[ \hat S_j'(\omega) \hat \rho \hat S_j'(\omega')^\dagger - \hat S_j'(\omega')^\dagger \hat S_j'(\omega) \hat \rho \right] + \hc,}\\
    \mathcolorbox{white}{C_{jk}(\omega)} &\mathcolorbox{white}{= \int_0^\infty e^{i\omega s}\langle \hat R_j^\dagger(t) \hat R_k(s-t) \rangle \dd s = \int_0^\infty e^{i\omega s}\langle \hat R_j^\dagger(s) \hat R_k(0) \rangle \dd s, }
\end{align}
\end{widetext}
where in the second line we used the stationarity of $\hat \rho_R$. Finally, we make the secular approximation and neglect rapidly rotating terms. Thus, we finally arrive at
\begin{align}
    \mathcolorbox{white}{\partial_t \hat \rho }& \mathcolorbox{white}{= \sum_{\omega, j,k}  C_{jk}(\omega) \left[ \hat S_j'(\omega) \hat \rho \hat S_k'(\omega)^\dagger - \hat S_k'(\omega)^\dagger \hat S_j'(\omega) \hat \rho \right] + \hc }
\end{align}
From here, we can define
\begin{align}
    \mathcolorbox{white}{\gamma_{jk}(\omega) }&\mathcolorbox{white}{= C_{jk}(\omega) + C_{kj}(\omega)^*, }\\
    \mathcolorbox{white}{h_{jk}(\omega) }&\mathcolorbox{white}{= \frac{-i}{2} \left( C_{jk}(\omega) - C_{kj}(\omega)^* \right),} \\
    \mathcolorbox{white}{\hat H_{LS} }&\mathcolorbox{white}{= \sum_{jk\omega} h_{jk}(\omega) \hat S_j'^\dagger(\omega) \hat S_k'(\omega).} \label{eqn:LSHam}
\end{align}
$\hat H_{LS}$ is the Lamb shift Hamiltonian, and $\gamma_{jk}$ describes the dissipative part of the dynamics:
\begin{align}
    \mathcolorbox{white}{\partial_t \hat \rho} &\mathcolorbox{white}{= -i[\hat H_{LS}, \hat \rho] + \L_{\mathrm{diss}} \hat \rho ,}\\
    \mathcolorbox{white}{\L_{\mathrm{diss}} \hat \rho} &\mathcolorbox{white}{= \sum_{\omega, j,k} \gamma_{jk}(\omega)  \left[ \hat S_j'(\omega) \hat \rho \hat S_k'(\omega)^\dagger - \frac{1}{2} \left\{ \hat S_k'(\omega)^\dagger \hat S_j'(\omega),  \hat \rho \right\} \right] .}
\end{align}
The matrix $\gamma_{jk}(\omega)$ is positive, and so it can be unitarily diagonalized. Taking $U(\omega)$ to be a unitary matrix, we can rewrite
\begin{align}
    \mathcolorbox{white}{\gamma_{jk}(\omega) }&\mathcolorbox{white}{ = \kappa_l(\omega) U_{lj}^*(\omega) U_{lk}(\omega).}
\end{align}
The dissipation can then be rewritten as
\begin{align}
    \mathcolorbox{white}{\L_{\mathrm{diss}} \hat \rho} & \mathcolorbox{white}{= \sum_{\omega, l} \D[\hat L_{l}(\omega)]\hat \rho,} \\
   \mathcolorbox{white}{\hat L_l(\omega)} &\mathcolorbox{white}{= \sum_n U_{ln}^*(\omega) \hat S_n'(\omega).} \label{eqn:finalJumpOps}
\end{align}
Recall that when we began, each operator $\hat S_j'$ was local to one of the two subsystems [c.f.~\cref{eqn:localSOPs}]. Moreover, because $\hat H_S = \hat H_A + \hat H_B$ contains no interactions between the subsystems, the projections into frequency space $\hat S_j'(\omega)$ must also remain local to a single sublattice. (I.e., the dynamics of a two non-interacting subsystems cannot mix local operators together). Thus, each jump operator in \cref{eqn:finalJumpOps} is a sum of local operators, as we set out to prove. Moreover, we can observe that the Lamb shift Hamiltonian in \cref{eqn:LSHam} is quadratic in $\hat S_j'(\omega)$, and so after tracing out the reservoir there will generically be long range Hamiltonian interactions between the two systems. Such a Lamb shift Hamiltonian is generically constrained by the form of the dissipation (via Kramers-Kronig relations); however, we take it to be arbitrary since it does not affect any of the final results given in the main text.

\section{Numerical Techniques}
\label{app:7}

In this section, we outline the numerical techniques used to generate the random models as presented in, e.g., \cref{fig:2}. The algorithm used can be summarized into the following steps:
\begin{enumerate}[1.]
    \item \textit{Generate the steady state}. The first step is generating the steady state. We pick a value of the entanglement $\delta \Erel$ and run an optimization algorithm to find a set of Schmidt coefficients such that the normalization is fixed to unity and the Renyi-2 entropy is fixed, beginning from a set of random values.
    \item \textit{Generate Jump Operators}. Next, we generate a random $N \times N$ matrix sampled from the random complex Ginibre ensemble which becomes the $A$ matrix. (If we wish to use a different ensemble, we follow the prescription in \cref{eqn:distributions}.) We then use this along with the Schmidt coefficients from Step 1 to generate $B$ uniquely using \cref{eqn:A-B-Relation}. Repeat this $M$ times to generate $M$ jump operators.
    \item \textit{Generate Hamiltonian}. If we are simulating directional dynamics, we generate a Hamiltonian as prescribed in \cref{eqn:CQA_Ham}. Otherwise, proceed to the next step.
    \item \textit{Find Dissipative Gap}. Given the matrices $A,B$ generated in Step 2 and Hamiltonian in Step 3 (if present), we use the QuTip \cite{Johansson2012} package to generate a Liouvillian superoperator. We extract its eigenvalues, from which we can find the dissipative gap.
\end{enumerate}
These steps generate a single data point for a fixed $N,M, \delta \Erel$, and distribution. Each plot samples from 20 different values of $\delta \Erel$, and samples different random models from each value to get good convergence of the average and standard deviation, usually about 100 samples for each.

To be even more explicit, we will now go through a specific example of generating such a system for $N = 2$ and $M=1$. First we perform Step 1. Note that because we have two constraints (normalization and entanglement), and two Schmidt coefficients, they are exactly specified. We can write the matrix $\Psi$ as
\begin{align}
    \mathcolorbox{white}{\Psi} & \mathcolorbox{white}{= \left(
    \begin{array}{cc}
        \frac{1}{2} - p & 0 \\
         0 & \frac{1}{2} + p
    \end{array}
    \right), }\\
    \mathcolorbox{white}{p} & \mathcolorbox{white}{= \sqrt{\frac{\delta \Erel}{2}}.}
\end{align}
Next, we do Step 2 and generate a random matrix $A$, which we will leave unspecified as
\begin{align}
    \mathcolorbox{white}{A} & \mathcolorbox{white}{= \left(
    \begin{array}{cc}
        a_{11} & a_{12} \\
        a_{21} & a_{22}
    \end{array}
    \right).}
\end{align}
This allows us to uniquely specify a matrix $B$ as
\begin{align}
    \mathcolorbox{white}{B} & \mathcolorbox{white}{= -\Psi A^T \Psi^{-1}= \left(
    \begin{array}{cc}
        -a_{11} &  \xi a_{21} \\
        \xi^{-1} a_{12} & -a_{22}
    \end{array}
    \right),} \\
    \mathcolorbox{white}{\xi} & \mathcolorbox{white}{= \frac{2p-1}{2p + 1}.}
\end{align}
Since we are not using directional dynamics, we can now proceed directly to step 4, and generate a Liouvillian superoperator. To do this, we first need to find the jump operator $L = A \otimes \1 + \1 \otimes B$, which can be written as
\begin{align}
    \mathcolorbox{white}{L} &\mathcolorbox{white}{ = \left(
    \begin{array}{cccc}
        0 & \xi a_{21} & a_{12} & 0 \\
        \xi^{-1} a_{12} & a_{11}-a_{22} & 0 & a_{12} \\
        a_{21} & 0 & a_{22} - a_{11} &  \xi a_{21} \\
        0 & a_{21} &  \xi^{-1} a_{12} & 0
    \end{array}
    \right).}
\end{align}
From here, it is simple to use this matrix to generate a Lindbladian and find its spectrum.
\end{document}